\documentclass[prd,aps,letterpaper,twocolumn,superscriptaddress,preprintnumbers,nofootinbib,floatfix]{revtex4}
\pdfoutput=1
\usepackage{color}
\usepackage[pdftex]{graphicx}
\usepackage{amsmath}
\usepackage{amsfonts}
\usepackage{amssymb}
\usepackage{rotating}
\usepackage{subfigure}
\usepackage{paralist} 
\usepackage{verbatim}
\usepackage{float}

\usepackage{soul} 
\usepackage{appendix}

\usepackage{natbib}
\usepackage{epsfig}
\usepackage[colorlinks=true,linkcolor=red,urlcolor=blue,citecolor=blue]{hyperref}

\usepackage[utf8]{inputenc}
\usepackage[vietnamese,english]{babel}
\usepackage{tabularx}
\newcolumntype{C}{>{\centering\arraybackslash}X}

\usepackage{mathpazo} 
\usepackage{tgpagella} 

\definecolor{ao(english)}{rgb}{0.0, 0.5, 0.0}

\newcommand{\ptettsimninety}{0.54}
\newcommand{\ptetesimninety}{0.17}
\newcommand{\pteeesimninety}{0.08}
\newcommand{\ptebbsimninety}{0.81}    
\newcommand{\ptettsimhundredfifty}{0.40}
\newcommand{\ptetesimhundredfifty}{0.12}
\newcommand{\pteeesimhundredfifty}{0.12}
\newcommand{\ptebbsimhundredfifty}{0.91} 
\newcommand{\ptettsimcross}{0.92}
\newcommand{\ptetesimcross}{0.47}
\newcommand{\pteetsimcross}{0.53}
\newcommand{\pteeesimcross}{0.60}
\newcommand{\ptebbsimcross}{0.73}

\newcommand{\snrttninety}{5.5 \sigma}
\newcommand{\snrteninety}{1.6 \sigma}
\newcommand{\snreeninety}{2.6 \sigma}
\newcommand{\snrbbninety}{0.6 \sigma}
\newcommand{\snrtthundredfifty}{8.7 \sigma}
\newcommand{\snrtehundredfifty}{2.6 \sigma}
\newcommand{\snreehundredfifty}{5.1 \sigma}
\newcommand{\snrbbhundredfifty}{2.4 \sigma }
\newcommand{\snrttcross}{4.3 \sigma}
\newcommand{\snrtecross}{- }
\newcommand{\snretcross}{2.6 \sigma}
\newcommand{\snreecross}{2.5 \sigma}
\newcommand{\snrbbcross}{0.6 \sigma}

\newcommand{\effttninety}{29\%}
\newcommand{\effteninety}{25\%}
\newcommand{\effeeninety}{29\%}
\newcommand{\effbbninety}{17\%}
\newcommand{\efftthundredfifty}{30\%}
\newcommand{\efftehundredfifty}{26\%}
\newcommand{\effeehundredfifty}{30\%}
\newcommand{\effbbhundredfifty}{20\%}
\newcommand{\effttcross}{40\%}
\newcommand{\efftecross}{0\%}
\newcommand{\effetcross}{63\%}
\newcommand{\effeecross}{47\%}
\newcommand{\effbbcross}{31\%}

\newcommand{\ptettdataninety}{0.17}
\newcommand{\ptetedataninety}{0.62}
\newcommand{\pteeedataninety}{0.58}
\newcommand{\ptebbdataninety}{0.04}
\newcommand{\ptettdatahundredfifty}{0.36}
\newcommand{\ptetedatahundredfifty}{0.96}
\newcommand{\pteeedatahundredfifty}{0.97}
\newcommand{\ptebbdatahundredfifty}{0.93}
\newcommand{\ptettdatacross}{0.41}
\newcommand{\ptetedatacross}{0.98}
\newcommand{\pteetdatacross}{0.43}
\newcommand{\pteeedatacross}{0.97}
\newcommand{\ptebbdatacross}{0.12}
\newcommand{\pteALL}{0.35}
\newcommand{\redChiALL}{1.02}

\newcommand{\hubbleConstraint}{{\ensuremath 67.3 \pm 3.5}~{\rm{km~s^{-1} Mpc^{-1}}} }
 
\newcommand{\sigmaEightConstraint}{{\ensuremath 0.819 \pm 0.034}}

\newcommand{\ptedustcrossPlanckNinty}{0.6}
\newcommand{\ptedustcrossPlanckHunderFifty}{0.4}

\newcommand{\snrparamshift}{3.3\sigma} 
\newcommand{\pteparamshift}{0.0008} 

\newcommand{\snrparamshiftLCDMBlock}{1.0\sigma} 
\newcommand{\pteparamshiftLCDMBlock}{0.31} 

\newcommand{\snrparamshiftFGBlock}{3.4\sigma} 
\newcommand{\pteparamshiftFGBlock}{0.0007}

\newcommand{\pteparamshiftKSZXI}{1.00}
\newcommand{\snrparamshiftKSZXI}{0.0\sigma}
 
\newcommand{\pteparamshiftExtraGalTempRest}{0.04}
\newcommand{\snrparamshiftExtraGalTempRest}{2.0\sigma}

\newcommand{\snrparamshiftExtraGalTemp}{2.7\sigma} 
\newcommand{\snrparamshiftGalPol}{2.6\sigma} 
\newcommand{\snrparamshiftGalPolExtraGalTemp}{3.7\sigma} 
\newcommand{\snrparamshiftFGRest}{0.5\sigma}

\newcommand{\WMAPpteparamshift}{0.18} 

\newcommand{\GalDustpteparamshift}{0.66} 

\newcommand{\Gsnrparamshift}{1.7\sigma} 
\newcommand{\Gpteparamshift}{0.092} 

\newcommand{\NGsnrparamshift}{1.9\sigma} 
\newcommand{\NGpteparamshift}{0.059}

\newcommand{\ExtraSmoothAsnrparamshift}{1.8\sigma} 
\newcommand{\ExtraSmoothApteparamshift}{0.076} 
\newcommand{\ExtraSmoothAsnrparamshiftLCDMBlock}{2.1\sigma}
\newcommand{\ExtraSmoothApteparamshiftLCDMBlock}{0.032} 
\newcommand{\ExtraSmoothAsnrparamshiftFGBlock}{1.5\sigma} 
\newcommand{\ExtraSmoothApteparamshiftFGBlock}{0.14}

\newcommand{\ExtraSmoothBsnrparamshift}{2.5\sigma} 
\newcommand{\ExtraSmoothBpteparamshift}{0.013} 
\newcommand{\ExtraSmoothBsnrparamshiftLCDMBlock}{2.8\sigma} 
\newcommand{\ExtraSmoothBpteparamshiftLCDMBlock}{0.0054}
\newcommand{\ExtraSmoothBsnrparamshiftFGBlock}{1.4\sigma} 
\newcommand{\ExtraSmoothBpteparamshiftFGBlock}{0.17}

\newcommand{\MeanExtraSmoothAsnrparamshiftLCDMBlock}{1.8\sigma} 
\newcommand{\MeanExtraSmoothApteparamshiftLCDMBlock}{0.076} 

\newcommand{\MeanExtraSmoothBsnrparamshiftLCDMBlock}{3.1\sigma} 
\newcommand{\MeanExtraSmoothBpteparamshiftLCDMBlock}{0.0023} 

\definecolor{orange}{rgb}{1,0.3,0}

\newcommand{\hubbleConstraintOrig}{{\ensuremath 64.6 \pm 3.3}~{\rm{km~s^{-1} Mpc^{-1}}} }
\newcommand{\sigmaEightConstraintOrig}{{\ensuremath 0.840 \pm 0.032}}
\newcommand{\snrparamshiftOrig}{2.7\sigma}
\newcommand{\pteparamshiftOrig}{0.0063}

\newcommand{\arizonaState}{School of Earth and Space Exploration, Arizona State University, Tempe, AZ, 85287, USA}
\newcommand{\bccp}{Berkeley Center for Cosmological Physics, Department of Physics,
University of California, Berkeley, CA 94720, USA}
\newcommand{\caltech}{Department of Physics, California Institute of Technology, Pasadena, California 91125, USA}
\newcommand{\cambridge}{Department of Applied Mathematics and Theoretical Physics, University of Cambridge, Wilberforce Road, Cambridge CB3 0WA, UK}
\newcommand{\cardiff}{School of Physics and Astronomy, Cardiff University, The Parade, 
Cardiff, Wales, UK CF24 3AA}
\newcommand{\cca}{Center for Computational Astrophysics, Flatiron Institute, 162 5th Avenue, New York, NY, USA 10010}
\newcommand{\columbia}{Department of Physics, Columbia University, New York, NY, USA 10027}
\newcommand{\cornellPhysics}{Department of Physics, Cornell University, Ithaca, NY, USA 14853}
\newcommand{\cornellAstro}{Department of Astronomy, Cornell University, Ithaca, NY, USA 14853}
\newcommand{\dunlapDepartment}{David A. Dunlap Department of Astronomy and Astrophysics, University of Toronto, 50 St. George Street, Toronto ON M5S3H4}
\newcommand{\dunlapInstitute}{Dunlap Institute for Astronomy and Astrophysics, University of Toronto, 50 St. George Street, Toronto ON M5S3H4}
\newcommand{\haverford}{Department of Physics and Astronomy, Haverford College, Haverford, PA, USA 19041}
\newcommand{\kavliCambridge}{Kavli Institute for Cosmology, University of Cambridge, Madingley Road, Cambridge CB3 OHA, UK}
\newcommand{\KwaZulu}{Astrophysics Research Centre, School of Mathematics, Statistics \& Computer Science, University of KwaZulu-Natal, Westville Campus, Durban 4041, South Africa}
\newcommand{\lbnl}{Lawrence Berkeley National Laboratory, One Cyclotron Road, Berkeley, CA 94720, USA}
\newcommand{\milano}{Department of Physics, University of Milano - Bicocca, Piazza della Scienza, 3 - 20126, Milano (MI), Italy}
\newcommand{\nasaGoddard}{NASA/Goddard Space Flight Center, Greenbelt, MD, 20771, USA}
\newcommand{\nist}{NIST Quantum Devices Group, 325
Broadway Mailcode 817.03, Boulder, CO, USA 80305}
\newcommand{\parisSaclay}{Laboratoire de l'Acc\'el\'erateur Lin\'eaire, Univ. Paris-Sud, CNRS/IN2P3, Universit\'e Paris-Saclay, Orsay, France}
\newcommand{\pennStateAstro}{Department of Astronomy and Astrophysics, The Pennsylvania State University, University Park, PA 16802, USA}
\newcommand{\pennStateInst}{Institute for Gravitation and the Cosmos, The Pennsylvania State University, University Park, PA 16802, USA}
\newcommand{\perimeter}{Centre for the Universe, Perimeter Institute for Theoretical Physics, Waterloo, ON, Canada N2L 2Y5}
\newcommand{\pitt}{Department of Physics and Astronomy, University of Pittsburgh, 
Pittsburgh, PA, USA 15260}
\newcommand{\princetonAstro}{Department of Astrophysical Sciences, Princeton University, 4 Ivy Lane, Princeton, NJ, USA 08544}
\newcommand{\princetonPhysics}{Joseph Henry Laboratories of Physics, Jadwin Hall, Princeton University, Princeton, NJ, USA 08544}
\newcommand{\rutgers}{Department of Physics and Astronomy, Rutgers, The State University of New Jersey, Piscataway, NJ USA 08854-8019}
\newcommand{\stonybrook}{Physics and Astronomy Department, Stony Brook University, Stony Brook, NY  11794}
\newcommand{\upenn}{Department of Physics and Astronomy, University of
Pennsylvania, 209 South 33rd Street, Philadelphia, PA, USA 19104}

\begin{document}
\title{The Atacama Cosmology Telescope: Delensed Power Spectra and Parameters}

\author{Dongwon~Han}
\affiliation{\stonybrook}
\affiliation{\cca}

\author{Neelima~Sehgal}
\affiliation{\stonybrook}
\affiliation{\cca}

\author{Amanda~MacInnis}
\affiliation{\stonybrook}

\author{Alexander van Engelen}
\affiliation{\arizonaState}

\author{Blake D. Sherwin}
\affiliation{\cambridge}
\affiliation{\kavliCambridge}

\author{Mathew S.~Madhavacheril}
\affiliation{\perimeter}
\affiliation{\princetonAstro}

\author{Simone Aiola}
\affiliation{\cca}

\author{Nicholas Battaglia}
\affiliation{\cornellAstro}

\author{James~A.~Beall}
\affiliation{\nist}

\author{Daniel~T.~Becker}
\affiliation{\nist}

\author{Erminia Calabrese}
\affiliation{\cardiff}

\author{Steve~K.~Choi}
\affiliation{\cornellPhysics}
\affiliation{\cornellAstro}

\author{Omar Darwish}
\affiliation{\cambridge}

\author{Edward~V.~Denison}
\affiliation{\nist}

\author{Mark~J.~Devlin}
\affiliation{\upenn}

\author{Jo~Dunkley}
\affiliation{\princetonAstro}
\affiliation{\princetonPhysics}

\author{Simone Ferraro}
\affiliation{\bccp}
\affiliation{\lbnl}

\author{Anna~E. Fox}
\affiliation{\nist}

\author{Matthew Hasselfield}
\affiliation{\cca}
\affiliation{\pennStateAstro}
\affiliation{\pennStateInst}

\author{J.~Colin Hill}
\affiliation{\columbia}
\affiliation{\cca}

\author{Gene~C.~Hilton}
\affiliation{\nist}

\author{Matt Hilton}
\affiliation{\KwaZulu}

\author{Ren\'ee~Hlo\v{z}ek}
\affiliation{\dunlapDepartment}
\affiliation{\dunlapInstitute}

\author{Johannes~Hubmayr}
\affiliation{\nist}

\author{John~P.~Hughes}
\affiliation{\rutgers}

\author{Arthur~Kosowsky}
\affiliation{\pitt}

\author{Jeff~Van~Lanen}
\affiliation{\nist}

\author{Thibaut~Louis}
\affiliation{\parisSaclay}

\author{Kavilan~Moodley}
\affiliation{\KwaZulu}

\author{Sigurd Naess}
\affiliation{\cca}

\author{Toshiya Namikawa}
\affiliation{\cambridge}

\author{Federico Nati}
\affiliation{\milano}

\author{John~P.~Nibarger}
\affiliation{\nist}

\author{Michael~D.~Niemack}
\affiliation{\cornellPhysics}
\affiliation{\cornellAstro}

\author{Lyman~A.~Page}
\affiliation{\princetonPhysics}

\author{Bruce Partridge}
\affiliation{\haverford}

\author{Frank J. Qu}
\affiliation{\cambridge}

\author{Alessandro Schillaci}
\affiliation{\caltech}

\author{David~N.~Spergel}
\affiliation{\cca}
\affiliation{\princetonAstro}

\author{Suzanne Staggs}
\affiliation{\princetonPhysics}

\author{Emilie~Storer}
\affiliation{\princetonPhysics}

\author{Edward J. Wollack}
\affiliation{\nasaGoddard}

\begin{abstract}
We present $\Lambda$CDM cosmological parameter constraints obtained from delensed microwave background power spectra.  Lensing maps from a subset of DR4 data from the Atacama Cosmology Telescope (ACT) are used to undo the lensing effect in ACT spectra observed at 150 and 98~GHz. At 150~GHz, we remove the lensing distortion with an effective efficiency of $\efftthundredfifty ~(TT )$, $\effeehundredfifty~(EE)$, $\efftehundredfifty~(TE)$ and  $\effbbhundredfifty~(BB)$; this results in detections of the delensing effect at $\snrtthundredfifty~(TT)$, $\snreehundredfifty~(EE)$, $\snrtehundredfifty~(TE)$, and $\snrbbhundredfifty~(BB)$ significance.  The combination of 150 and 98~GHz $TT, EE,$ and $TE$ delensed spectra is well fit by a standard $\Lambda $CDM model.  We also measure the shift in best-fit parameters when fitting delensed versus lensed spectra; while this shift does not inform our ability to measure cosmological parameters, it does provide a three-way consistency check among the lensing inferred from the best-fit parameters, the lensing in the CMB power spectrum, and the reconstructed lensing map. This shift is predicted to be zero when fitting with the correct model since both lensed and delensed spectra originate from the same region of sky. Fitting with a $\Lambda $CDM model and marginalizing over foregrounds, we find that the shift in cosmological parameters is consistent with zero.  Our results show that gravitational lensing of the microwave background is internally consistent within the framework of the standard cosmological model.
\end{abstract}

\maketitle

\section{Introduction}
\label{sec:intro}
\setcounter{footnote}{0} 
Measurements of the cosmic microwave background (CMB) power spectra have yielded powerful constraints on cosmological parameters~\cite[e.g.][]{Planck2018Parameters, Hinshaw2013, Henning2018, Choi2020, Aiola2020}.  Gravitational lensing of the microwave background distorts these power spectra; lensing smooths the acoustic peaks of temperature and E-mode power spectra and generates B-mode power spectra~\cite{Blanchard1987,Bernardeau1997,Lewis2006,Zaldarriaga1998}.  Since the amplitude of gravitational lensing fluctuates from one patch of the Universe to another patch, lensing induces additional correlations in the CMB angular power spectrum.  Delensing the maps undoes both effects and reduces the uncertainty in the resulting band powers.

Delensing was first proposed as a means to remove lens-induced signal on large-scale B modes to measure the imprint of primordial gravitational waves \cite{Knox2002,Kesden2002,Seljak2004,Smith2012}.  At the same time, it has also been shown that delensing small-scale temperature or polarization would give a cleaner view of the last-scattering surface and help constrain some of the physics in the acoustic peaks and diffusion damping in that regime.  In particular,~\cite{Green2017} showed that extra relativistic species present in the early Universe could be better constrained with the improved acoustic peak localization that delensing makes possible. Delensing at small scales has been demonstrated with observational data on several occasions, including with external tracers of the lensing field~\cite{Larsen2016,Manzotti2017, Planck2018Lensing} as well as for lensing maps obtained internally with the CMB~\cite{Carron2017, Planck2018Lensing, Adachi2019}.

The delensing procedure consists of constructing an estimate of the specific realization of the matter distribution responsible for the lensing of the CMB; this estimated mass map is then used to remap points in the CMB map to their original undeflected positions. Delensing is predicted to increase the extracted cosmological information by tightening parameter constraints when combining primordial CMB and CMB lensing power spectra measurements;  this is because delensing removes the uncertainty in the realization of the intervening lensing matter distribution~\cite{Green2017}. 

In this work, we present the first $\Lambda$CDM parameter constraints obtained from delensed power spectra. Since CMB lensing power spectrum measurements are not included, these parameter constraints are not expected to be tighter than those from the lensed CMB spectra. We also measure the shift in best-fit parameters when fitting delensed versus lensed CMB spectra. When fitting with the correct model, we expect no detectable shift in parameters; in addition, the uncertainty on this shift does not suffer from sample variance since both the lensed and delensed spectra are sourced from the same region of sky. 

The parameter-shift statistic we introduce in this work provides a three-way consistency check among the lensing inferred from the best-fit parameters, the lensing in the CMB power spectrum, and the reconstructed lensing map. In general, 
an inconsistency among these three yields a shift between
the $\Lambda$CDM parameters inferred from the lensed and delensed power spectra. This consistency test can be used to explore inaccurate modeling of foregrounds or secondaries, systematic errors, and departures from $\Lambda$CDM.  While the subset of Atacama Cosmology Telescope (ACT) data used in this work does not yet have the significance to weigh in on potential new physics in the early Universe that could, for example, probe the recently noted tension between low and high-redshift values of $H_0$~\cite{Riess2018, Verde2019,Knox2019}, with future CMB data sets, we can use this parameter-shift statistic
to probe these types of models. In particular, this parameter-shift test provides a novel way to search for new physics or systematic effects that may be degenerate with lensing-induced peak smoothing.  Examples of models with new physics that can mimic lensing-induced peak smoothing are discussed in~\cite{Hazra2014,Munoz2016,Smith2017,Hazra2019,Knox2019,Domenech2019,Domenech2020}. We discuss our current sensitivity to inconsistent lensing in more detail in Section~\ref{sec:newPhysics}, using simulations that match the properties of the subset of ACT data used in this work.

In section~\ref{sec:data} and~\ref{sec:sims}, we discuss the ACT data used in this work, and the simulations that model this data.  The delensing pipeline is presented in section~\ref{sec:pipe}, including our procedure to remove a known bias that arises when delensing CMB maps with reconstructions of the lensing potential obtained from those same maps~\cite{Teng2011, Namikawa2014, Sehgal2017, Carron2017, Namikawa2017, Planck2018Lensing, Adachi2019, Baleato2020}.  We present the delensed power spectra in section~\ref{sec:spec}, 
and the likelihood developed for delensed spectra in section~\ref{sec:like}. Section~\ref{sec:params} shows the resulting cosmological parameters, followed by a discussion in section~\ref{sec:discussion}.  The analysis products presented in this work are public as part of the ACT DR4 data release on the NASA Legacy Archive for Microwave Background Data Analysis.{\color{blue}\footnote{https://lambda.gsfc.nasa.gov/product/act/}}

\section{Data}
\label{sec:data}

We analyze ACT data collected from two seasons of observations.  In particular, this analysis focuses on one region of the sky, labeled D56, which spans 565 square degrees (about $1\%$ of the full sky).  The coordinates and map noise levels are given in a companion paper~\cite{Choi2020} (hereafter C20). D56 was observed in both the 2014 and 2015 seasons (hereafter s14 and s15).

The Atacama Cosmology Telescope Polarimeter (ACTPol)~receivers consist of three bolometer arrays sensitive to both temperature and polarization~\cite{Thornton2016}.  From s14 through s15, data were obtained at 150 GHz.  In s15, 98 GHz data were also obtained.  All data used in this analysis were taken during the nighttime to minimize beam and pointing variations induced by solar heating of the telescope.  Details of the mapmaking process, a variety of null tests performed on these maps, and their corresponding spectra are discussed in~C20 and a companion paper~\cite{Aiola2020} (hereafter A20).  Four map splits are made by separating the time-ordered data into equal time splits for each combination of season (s14, s15), detector array (PA1, PA2, PA3), frequency (98, 150 GHz), and temperature/polarization type ($T, Q, U$) as described in~A20. Each map split is calibrated using the calibration factors described in~C20. Also, as described in~A20, point sources detected above $5\sigma$ at 150 GHz in a matched-filtered map are subtracted from each map split.  This roughly corresponds to removing all sources above 5~mJy from the map splits.{\color{blue}\footnote{Note that removing these sources does not result in a uniform flux threshold since the noise levels spatially vary in the matched-filtered map. Thus we later add back sources below 15 mJy to the maps used to make the power spectra in order to obtain a uniform flux threshold.}}

The power spectrum analysis in~C20 keeps the four map splits separate in order to maximize signal-to-noise ratio by taking as many cross spectra as possible to minimize the noise. Here, we coadd two sets of two splits (out of the four) to make two ‘coadded data splits’. These coadded splits are used to obtain both lensed and delensed power spectra.  Since the optimal delensing procedure involves delensing only signal-dominated modes~\cite{Green2017}, we do not expect a gain in delensing signal-to-noise ratio by making as many splits as in~C20; however, we still require some splits in order to reduce the noise bias. Furthermore, we coadd each set of four map splits to make one ‘coadded data map’ to use for the lensing reconstruction.

To do the coadding, each of the four map splits is first multiplied, pixel by pixel, by its corresponding inverse-variance weight map.  The products are then summed over pixel-wise to make the single coadded data map, $D_{coadd}$, and the two coadded data splits, $D_{coadd}^{split}$: 
\begin{equation}
D_{coadd} = ({\sum}_{i=1}^{4} W_i~D^{split_i}) /  {\sum}_{i=1}^{4} W_i,
\end{equation}
\begin{equation}
D_{coadd}^{split_j} = ({\sum}_{k} W_k~D^{split_k}) /  {\sum}_{k} W_k, 
\end{equation}
where $k \in [1,2]$ or $k \in [3,4]$.  Coadded weight maps, $W_{coadd}$, are made by averaging over the weight maps pixel-wise, i.e.~$W_{coadd} = {\sum}_{i=1}^{4} W_i / 4$ and $W_{coadd}^{split_j} = {\sum}_{k} W_k / 2$ where $k \in [1,2]$ or $k \in [3,4]$.

We next combine these individual season and array maps. First, we combine data from all seasons and arrays for the observations of patch D56, constructing a `patch map' of the region and two `patch-map splits'.  
The former is used for lensing reconstruction and the latter for obtaining the CMB power spectra, as mentioned above.  For the maps used in the lensing reconstruction, we also combine maps of different frequencies. For the maps used in calculating the CMB power spectra, we add back all detected and removed point sources below 15~mJy to each map to have a single flux threshold, as is done in~C20.

We make the above combinations by first convolving each map to a common beam. We choose the 98 GHz PA3 beam from season s15 to be the effective beam for this patch, since it is the largest beam; we deconvolve with the effective beam later.  All the coadded data maps and splits are convolved with the ratio of the effective beam to its original beam.  These coadded maps are then combined, weighted by the corresponding $W_{coadd}$:
\begin{equation}
D_{patch}^X = ({\sum}_{m} W_{coadd_m}^X ~ D_{coadd_m}^X) /  {\sum}_{m} W_{coadd_m}^X
\end{equation}
where $m$ sums over all seasons and arrays (and frequencies if relevant), and $X \in (\rm{map}, \rm{split_1}, \rm{split_2})$; this creates the patch map and patch-map splits. 

We choose the D56 `common boundary mask' described in~C20 as the analysis mask. 
(Note that using the `spatial window function mask' in~C20, which is spatially varying because it includes the weighting of the inverse noise variation, would distort the local gradient of the reconstructed lensing potential.)  We then multiply the patch map and patch-map splits with the analysis mask.  This results in an effective sky area of 482 square degrees.  In order to obtain the two-dimensional noise power spectrum, $N_{2D}$, of the patch map, for use in the lensing reconstruction filter, we subtract the cross spectra of the two map splits from the mean of their auto spectra. To correct for the analysis mask and account for the factor of two higher noise power in the splits compared to the full map, we normalize $N_{2D}$ by the number $1/(2w)$, where $w={\sum}_{i} (M_i)^2/{\sum}_{i} 1$, and $i$ sums over all pixels in the mask $M$.

Finally, we in-paint the temperature maps used in the lensing reconstruction at the positions of galaxy cluster candidates detected above $5\sigma$ at 150 GHz, using the method described in~\cite{Madhavacheril2019}. Similarly, we in-paint using a 5 arcminute radius both the temperature maps and the polarization maps at the {\it{positions}} of irregular-shaped sources or bright sources (detected with signal-to-noise ratio greater than 90).  In the D56 analysis area used here, we have seven irregular-shaped or bright sources.  Note that these sources have already been removed from the maps, but we are in addition in-painting a large radius around these sources in case there is any leakage from them into a surrounding area. In order to remove the large-scale ground contamination in our maps, we apply a Fourier-space mask that cuts out $\ell$ modes in the ranges $-90<\ell_{x}< 90$~and $-50<\ell_{y}<50$.  We then deconvolve the patch map and patch-map splits with the effective beam. We also compute the pixel window function in two-dimensional Fourier space, and deconvolve each map by this function as is done in~C20. We remove all $\ell>10,000$ modes from the maps for ease of analysis, since they contribute negligible signal-to-noise.  We refer to the resulting patch map and patch-map splits as {\it{prepared maps}}.  In total, we have three patch maps ($T, Q, U$) and 12 patch-map splits (two splits for $T, Q$ and $U$ at 150 GHz and 98 GHz).

\section{Simulations}
\label{sec:sims}

We make simulations of our full data set in order to verify our delensing pipeline, estimate biases to our delensed spectra, and obtain the covariance matrix for our spectra. In particular, we make simulated maps for each D56 season, array, and frequency using the simulation software pipeline described in~C20. The cosmology we use for these simulations is based on {\it{Planck2015}} parameters.{\color{blue}\footnote{These {\it{Planck2015}}-based simulations use $\Omega_b h^2=0.02219$, $\Omega_c h^2=0.1203$, $h=0.6702$, optical depth $\tau=0.066$, amplitude of scalar perturbations $A_s=2.151\times 10^{-9}$, and scalar spectral index of $n_s=0.9625$. We take $k_0 = 0.05$ Mpc$^{-1}$ as the pivot scale and the total mass of neutrinos as 0.06 eV.}} We construct these simulations to include Gaussian foregrounds in the temperature maps matched to the levels measured in the data. We do not model any polarization foregrounds in these simulations (see more discussion in Section~\ref{sec:results}).  These are the simulations used throughout this work unless otherwise stated.

For the maps we use to calculate the CMB power spectra, where we add back all detected point sources below 15 mJy to each map as discussed above, we use the foreground power spectrum templates described in~C20 to construct Gaussian temperature foreground realizations.  Since the maps we use to make the lensing reconstructions have lower foreground levels to minimize foreground-induced bias, we generate separate foreground templates for them from the data. We do this by first taking the cross spectra of the data splits to obtain spectra at $150 \times 150$ GHz, $98  \times 150$ GHz, and $98 \times 98$ GHz.  In order to fit our high-$\ell$ foreground model to the data in the $\ell$ range of [2000, 8000], we first subtract both CMB and thermal Sunyaev-Zel'dovich (tSZ) effect theory power spectra calculated using the publicly available {\it{szar}} package~\cite{Madhavacheril2017}. This is done in order to remove an approximation to the lower-$\ell$ contributions of the data, while still remaining blind to the actual low-$\ell$ data spectra. We fit the residual spectra to a polynomial function to capture the contributions of the kinetic Sunyaev-Zel'dovich effect, radio galaxies, and dusty star-forming galaxies. Finally, we add back the tSZ theory curve to the fitted spectra.  We use these foreground power spectrum templates to make Gaussian temperature foreground realizations.  As discussed later in Section~\ref{sec:FG}, we investigate the impact of switching to more realistic simulations including non-Gaussian foregrounds; these more realistic foregrounds also have correlations between the foregrounds and the lensing potential itself. 

We follow the procedure described in Section~\ref{sec:data} to make simulated prepared patch maps and patch-map splits. In total, we make three sets of 512 simulated prepared maps ($set_1$, $set_2$, and $set_3$). $Set_1$ and $set_2$ share common lensing potential maps, but have independent realizations of the primordial CMB. $Set_3$ has independent realizations of both the primordial CMB and the lensing potential.  We use $set_1$ and $set_2$ to estimate the delensing bias and Monte Carlo (MC) bias discussed in Section~\ref{sec:pipe}; we use $set_3$ to verify the delensing pipeline, as we show in Figure~\ref{fig:verification}, and to obtain the covariance matrix.

\section{Delensing Pipeline}
\label{sec:pipe}

The steps of the delensing pipeline are to (1)~make a minimum-variance lensing reconstruction for the D56 patch, (2)~delens each $T, Q, U$ patch-map split using the lensing reconstruction, (3)~convert the delensed map splits to $TT, TE, EE$, and $BB$ power spectra, and (4)~subtract off a `delensing bias' and an `MC bias' from these spectra to obtain the final delensed power spectra.  We describe each of these steps in more detail below.

\subsection{Reconstructing Lensing Convergence Maps}
\label{sec:reconstruction}
We reconstruct the minimum-variance lensing convergence map for the D56 patch by first converting the $Q$ and $U$ prepared patch maps to $E$ and $B$ maps using the flat-sky pure $E\mbox{-}B$ decomposition method outlined in~\cite{Louis2013} and discussed first in~\cite{Smith2006}.  This $E\mbox{-}B$ decomposition has the advantage that it clearly isolates the low-cosmic-variance $B$-mode field, allowing us to mitigate and test for $E\mbox{-}B$ leakage; it also simplifies the analysis because $E$ and $B$ fields are independent of the coordinate system used.  We then generate two-dimensional filters used to make the lensing reconstructions by adding together the two-dimensional lensed CMB theory spectrum and foreground theory spectrum (assuming azimuthal symmetry), and two-dimensional noise power spectrum ($N_{2D}$ described in Section~\ref{sec:data}), appropriate for each map, after deconvolving the latter with the effective beam. In addition, we apply the following $\ell$-space cuts to the lensing filters:~($|\ell_{x_{\rm{min}}}|$, $|\ell_{y_{\rm{min}}}|$, $\ell_{\rm{min}}$, $\ell_{\rm{max}}$) = ($90, 50, 500, 3000$) for both temperature and polarization maps.  We apply the $|\ell_{x_{\rm{min}}}|$ and $|\ell_{y_{\rm{min}}}|$ cuts to remove ground pick up, the $\ell_{\rm{min}}$ cut to minimize atmospheric contamination and large-scale systematics, and the $\ell_{\rm{max}}$ cut to minimize extragalactic foreground contamination.  

Using these lensing filters and the $T, E,$ and $B$ maps, we reconstruct lensing convergence maps using a flat-sky, quadratic lensing estimator, following the procedure detailed in~\cite{Zaldarriaga1999,Hu2002,Sherwin2017}.{\color{blue}\footnote{Given the noise level in our maps, this quadratic estimator is optimal. For lower noise levels, one can improve the delensing efficiency by switching to an iterative maximum likelihood estimator~\cite{Smith2012}.}}  Our use of a flat-sky reconstruction algorithm is justified since the D56 sky patch is relatively small (about $1\%$ of the total sky area), close to the equator, and only spans about ten degrees in declination. We find a mean normalization bias across Fourier modes in the lensing reconstruction of $0.1\%$ due to the use of a flat-sky, as opposed to a curved-sky, reconstruction algorithm; we correct for this normalization bias with the MC bias correction discussed below.  We note that all other harmonic-space algorithmic steps throughout this work are done with spherical harmonic transforms, with the exception of the FFT-based lensing routine detailed in~\cite{Hu2002,Sherwin2017} and described above, the pixel window deconvolution, and the Fourier-space filtering. 

We further process the reconstructed lensing convergence maps, $\kappa$, by applying an $L$-space cut to each convergence map such that ($L_{\rm{min}}$, $L_{\rm{max}}$) = (80, 3000); the $L_{\rm{min}}$ cut restricts $\kappa$ to modes minimally impacted by uncertainty in the mean field map discussed below, and the $L_{\rm{max}}$ cut ensures we only include modes accurately modeled in the simulations.  Then we construct a two-dimensional estimate of the lensing noise power, $N(\mathbf{L})$, that is computed using Eq.~11 in~\cite{Hu2002}, which takes as input the $\ell$-space lensing filters described above. We construct the Fourier-space minimum-variance convergence map,
\begin{equation}
\kappa_{\rm{MV}}( \mathbf{L}) = N( \mathbf{L})_{MV}~(\sum_i \kappa( \mathbf{L})_i / N( \mathbf{L})_i),
\end{equation}
where~$N(\mathbf{L})_{MV}=(\sum_i 1/N(\mathbf{L})_i)^{-1}$ and~$i \in (TT, EE, EB)$.{\color{blue}\footnote{By excluding the $TE$ lensing estimator, the delensing bias in the $TE$ spectra becomes negligible. This is because we are not delensing this spectra using a lensing potential derived from the same $TE$ combination.  The absence of the $TE$ delensing bias provides a useful consistency test (see Section~\ref{sec:Clbias} and Figure~\ref{fig:verification}).}} For large-scale lensing modes (i.e $L < 150$), which are responsible for most of the delensing signal-to-noise ratio, the $TT, EE$ and $EB$ estimators contribute about $50\%$, $35\%$, and $15\%$, respectively, to $\kappa_{\rm{MV}}$.  For simulation $set_1$ and $set_3$ described in Section~{\ref{sec:sims}}, we repeat the steps above to reconstruct the $\kappa_{\rm{MV}}$ for each simulation and obtain the mean of all these $\kappa_{\rm{MV}}$ maps, which we call the mean field convergence map.{\color{blue}\footnote{Note that the mean field convergence map estimate does not include simulations from $set_2$ because we never make lensing reconstructions using $set_2$.}} We subtract this mean field map from each $\kappa_{\rm{MV}}$ map (data and all simulations) in order to remove the largest effect of the mode coupling due to the mask. 

Finally, we Wiener filter each $\kappa_{\rm{MV}}$ to downweight the noisy modes.  The Wiener-filtered convergence map is given by the Fourier transform of 
\begin{equation}
\kappa( \mathbf{L})_{\rm{MV}}^{W} = \kappa( \mathbf{L})_{\rm{MV}}~\frac{C^{\kappa\kappa}( \mathbf{L})^{theory}}{( C^{\kappa\kappa}( \mathbf{L})^{theory} + N( \mathbf{L})_{\rm{MV}})}, 
\end{equation}
where $C^{\kappa\kappa}( \mathbf{L})^{theory}$ is the theory convergence power spectrum used to generate the simulations described in Section~\ref{sec:sims}.  Thus we obtain final minimum-variance reconstructed convergence maps in Fourier space.{\color{blue}\footnote{Note that the D56 convergence map in~\cite{Darwish2020} differs from the one presented here in that it includes the $TE$ lensing estimator, and was constructed using CMB maps that were foreground cleaned following the method in~\cite{Madhavacheril2019} and coadded in Fourier space as opposed to real space.  When the $TE$ estimator is removed from the minimum-variance convergence map in~\cite{Darwish2020}, and we use the latter to delens the CMB maps described herein, then the delensed spectra are consistent, agreeing, for example, to within 20\% of the error bars for $TT$ at 150 GHz out to $\ell=4000$.}}

\subsection{Delensing Maps and Obtaining Power Spectra}
\label{sec:delensing}
We convert the final convergence maps, $\kappa_{\rm{MV}}^{W}( \mathbf{L})$, into lensing potential maps, $\phi_{\rm{MV}}( \mathbf{L})$, using the relation $\phi_{\rm{MV}}( \mathbf{L}) = 2 \kappa_{\rm{MV}}^{W}( \mathbf{L}) / L^2$ (obtained from converting the relation $\nabla^2 \phi_{\rm{MV}}= -2 \kappa_{\rm{MV}}^{W}$~\cite{Hu2007} into Fourier space). We then inverse Fourier transform each resulting $\phi_{\rm{MV}}( \mathbf{L})$ to a real-space lensing potential map.
Then we "lens" each of the two prepared map splits for $T, Q,$ and $U$ with the negative of the real-space potential map $(-\phi_{\rm{MV}})$ in order to delens them.{\color{blue}\footnote{This is technically anti-lensing, where the lensing potential is evaluated at the lensed, as opposed to the unlensed, position.  This is a good approximation to inverse-lensing, using the unlensed position for evaluating $\phi$~\cite{Green2017, Larsen2016, Anderes2015}.}}  The lensing algorithm used for this is the publicly available, flat-sky, {\it Taylens} software~\cite{Louis2013}.

We then multiply the lensed and delensed $T$, $Q$, and $U$ splits with a mask that has 5 and 8 arcminute radius holes for 150 and 98 GHz respectively, at the location of each point source that was subtracted from the maps; this removes the impact on the spectra of imperfect source subtraction.  We convert the $Q$ and $U$ maps to $E$ and $B$ maps by rotating in harmonic space using the curved-sky routines in the {\it{libsharp}} library~\cite{Reinecke2013}; since we undo mask-induced mode-coupling with the pseudo-$C_\ell$ formalism described below we do not need to use a pure $E\mbox{-}B$ decomposition as discussed in Section~\ref{sec:reconstruction}.

We calculate cross spectra between map splits for both the lensed and delensed maps using a curved-sky power spectrum routine that implements a standard pseudo-$C_\ell$ formalism; this Power-spectrum In Tracts Algorithm on the Sphere ($\it{PITAS}$) is publicly available at \url{https://github.com/dwhan89/pitas}.  
For this analysis, all the spectra are binned in the same manner as in~C20, using the $\ell$ range of $575 < \ell < 7925$ for $TT$ and $475 < \ell < 7925$ for $TE,EE,$ and, $BB$.  The $TT$ $\ell$ range matches that used in~C20 and~A20, and the minimum polarization $\ell$ of 475, which has been standard in prior ACT lensing analyses, is higher than what is used in C20/A20, which have a minimum $\ell$ of 325.  This gives us 47 spectral bins for $TT$ and 49 for $TE, EE,$ and $BB$.  We have verified that our power spectrum code gives results consistent with the power spectrum code used in~C20. The mean power spectrum over 500 simulations agrees to better than $0.6\%$ across all $\ell$s, which is an agreement within roughly $4\%$ of the error bars for a single simulation.  The error bars differ by up to $11\%$ before accounting for the difference in mask area and coadding procedure used; after correcting for the effective mask area, the error bars agree to better than $5\%$.

To correct for the missing Fourier modes that are removed by the Fourier-space mask discussed in Section~\ref{sec:data}, we compute a Fourier-space transfer function.  We calculate this transfer function by taking the ratio of the power spectra computed with and without the Fourier-space mask, and find the average of this over 512 simulations. We then correct the power spectra by this transfer function, which is $\ell$-dependent and ranges from a roughly $1\%$ departure from unity at high $\ell$ to a roughly $20\%$ departure from unity at the lowest $\ell$ we consider, $\ell=475$. 
We also compute the Fourier-space transfer function covariance matrix, $\Sigma_{kt}$, from these 512 transfer functions, and add it to both the lensed and delensed covariance matrices ($\Sigma_{len}$ and $\Sigma_{delen}$) described in the next sub-section.  

For the data power spectra, we make two additional corrections: (1)~we correct for aberration and modulation, and (2)~we apply a one-dimensional mapper transfer function. Aberration and modulation both arise from the motion of the Earth in the CMB rest frame; aberration is a frequency-independent geometric effect that looks like lensing, and modulation is a frequency-dependent intensity shift~\cite{Planck2013Doppler}. The aberration and modulation correction factors are computed by taking the mean difference in 512 lensed simulation spectra, before and after aberration and modulation are applied. Following~C20, we also compute one-dimensional mapper transfer functions by comparing spectra of simulations before and after being run through the ACT mapping pipeline described in~C20. This transfer function corrects for imperfections in the mapmaking pipeline, and is less than $0.5\%$ of the total power at all $\ell$s used in this analysis.  These two data-specific corrections are applied to both the lensed and delensed data spectra.\\
\\
\noindent We call the resulting delensed spectra for $TT, TE, EE,$ and $BB$ the {\it{raw delensed spectra}}. 

\subsection{Subtracting Delensing Bias and MC Bias}
\label{sec:subBias}
The raw delensed spectra are biased because the noise in the reconstructed $\phi_{\rm{MV}}$ map is correlated with the CMB map when delensing, as pointed out in~\cite{Teng2011, Namikawa2014, Sehgal2017, Carron2017, Namikawa2017, Planck2018Lensing, Adachi2019, Baleato2020}.  We estimate this bias, $C_\ell^{bias}$, as described below and subtract it from the raw delensed spectra, $C_\ell^{raw, delen}$.  In addition, we also subtract a Monte Carlo (MC) bias, $C_\ell^{mc}$, in order to account for small imperfections in the ability of the simulations to capture theory.  After these subtractions, both detailed below, we obtain the final delensed spectra, $C_\ell^{delen}$.   

To calculate $C_\ell^{bias}$ we use the two simulation sets, $set_1$ and $set_2$, that share common $\phi$ maps. We delens the prepared map splits from both $set_1$ and $set_2$ using the reconstructed $\phi_{\rm{MV_1}}$ from $set_1$. Then we calculate the raw delensed spectra of both sets, $C_\ell^{raw, delen_{S_1}}$ and $C_\ell^{raw, delen_{S_2}}$. Since we use $\phi_{\rm{MV_1}}$ to delens $set_{\rm{2}}$, then $C_\ell^{raw, delen_{S_2}}$ has no delensing bias, i.e.~$C_\ell^{raw, delen_{S_2}} = C_\ell^{delen_{S_2}}$.  

\begin{figure*}[t]
  \centering
  \hspace{-3mm}\includegraphics[width=\textwidth]{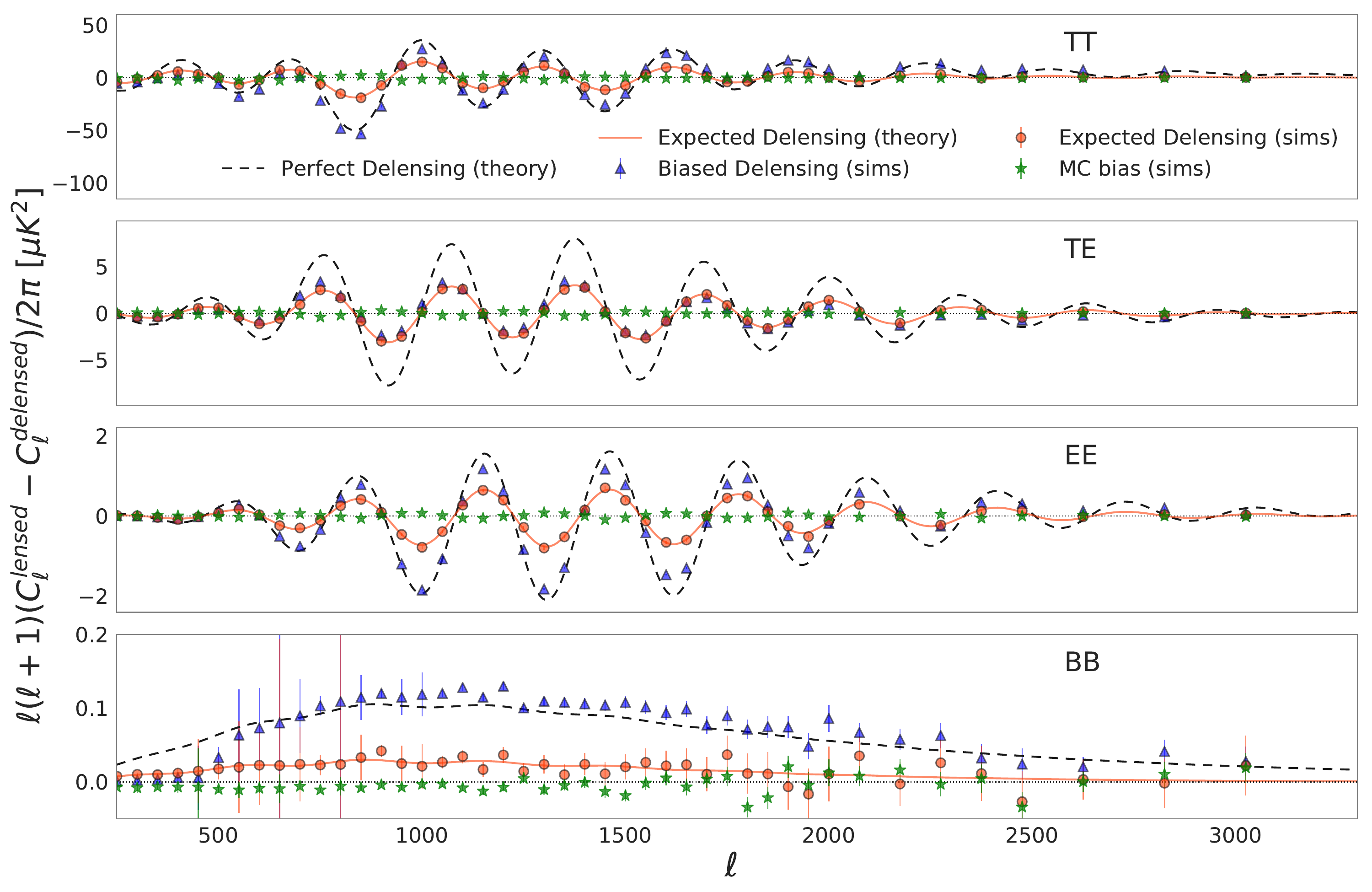}
  \caption{Shown are lensed minus delensed spectra from the mean of 512 simulations (described in Section~\ref{sec:sims}) for $150 \times 150$ GHz. The black dashed curves show ``perfect delensing'' assuming no noise, calculated from CAMB with the CMB theory spectra used to generate the simulations. The orange solid curves show the expected delensing achievable given ACT noise levels, also calculated from CAMB using the same CMB theory spectra along with the lensing noise power spectrum discussed in Section~\ref{sec:pipe}. The delensing procedure described in Sections~\ref{sec:pipe} and~\ref{sec:spec} is used to delens simulated ACT data from simulation $set_3$. The orange points show the delensed spectra after correcting for the delensing bias (obtained from simulation $set_1$ and $set_2$) and the MC bias (obtained from simulation $set_2$ and shown as green stars); these biases are discussed in Section~\ref{sec:pipe}.  These difference spectra clearly show the peak sharpening expected from delensing. The blue triangles show the delensed spectra prior to subtracting both the delensing bias and the MC bias.  Note that the fact that the amplitudes of the biased delensing simulations for $TT$, $EE$ and $BB$ are similar to the amplitudes of the perfect delensing is an artifact of the signal to noise of the ACT D56 data set.  By contrast, as described in the text, the $TE$ simulations have a negligible bias.  We also show the error on the mean of 512 simulations.}
  \label{fig:verification}
  \vspace{2mm}
\end{figure*}

Naively defining the delensing bias as the difference between $C_\ell^{raw, delen_{S_1}}$ and $C_\ell^{delen_{S_2}}$ does not cancel sample variance due to the different CMB background realizations for $set_{\rm{1}}$ and $set_{\rm{2}}$.  In order to cancel sample variance and reduce the error on $C_\ell^{bias}$, we calculate 
\begin{equation}
C_\ell^{{\mathit{bias}}} = \frac{\sum_i [ (C_{\ell_i}^{{\mathit{raw, delen, {s_1}}}} - C_{\ell_i}^{len, {s_1}} ) - (C_{\ell_i}^{delen, {s_2}} - C_{\ell_i}^{len, {s_2}}) ]}{N_{sim}}, 
\end{equation}
\\
where $C_\ell^{len}$ is the lensed spectra prior to delensing, $i$ is the simulation number, and $N_{sim}$ is the total number of simulations used to estimate the bias.  To get a sense of the size of $C_\ell^{bias}$, it is the difference between the biased delensing points (blue triangles) and the corrected delensing points (orange circles) in Figure~\ref{fig:verification}. Note that for the $TE$ spectra there is negligible delensing bias because we do not include the $TE$ lensing quadratic estimator in the minimum-variance lensing reconstruction.

To estimate and correct for several non-idealities in our analysis that are captured in our full simulation suite, we also compute a bias based on Monte Carlo simulations, i.e.~the MC bias.  This bias corrects for (1)~the small bias in $\phi_{\rm{MV}}$ from using a flat-sky code as discussed above, (2)~imperfect delensing around the edges of the patches, (3)~the slight change in mask-induced mode-coupling due to delensing the masked CMB map, and (4)~any change to the foreground power spectra due to the delensing procedure. This MC bias is calculated by 
\begin{equation}
C_\ell^{mc} = \frac{\sum_i [(C_{\ell_i}^{delen, {s_2}} - C_{\ell_i}^{len, {s_2}}) - (C_\ell^{delen, th} - C_\ell^{len, th})]}{N_{sim}}, 
\end{equation}
\\
where $C_\ell^{delen, th}$ and $C_\ell^{len, th}$ are theoretical delensed and lensed spectra calculated with the CAMB software package~\cite{Lewis2000} using the same parameters used to generate the simulations; the calculation of $C_\ell^{delen, th}$ is described in detail in Section~\ref{sec:like}.  We use simulation $set_2$ to calculate the MC bias since the delensed spectra from $set_2$ have no delensing bias.  We show $C_\ell^{mc}$ in Figure~\ref{fig:verification} (green stars), and note that it is a small change to the spectra.{\color{blue}{\footnote{In particular, the MC bias is less than a $0.4\%$ change to the $TT$ spectra for the full $\ell$ range, and less than a $1.7\%$ change to the $EE$ spectra below $\ell=3000$ at 150 GHz.}}}    

In addition, to account for the fact that the CMB power in the simulations is cut above $\ell=5100${\color{blue}\footnote{The lensed CMB part of the simulations used in this work was created with an $\ell=5100$ cut in order to save disk storage space since 512x3x3=4,608 CMB simulated maps were required for this analysis (the last factor of 3 is for the separate $T,Q$, and $U$ maps).}}, we apply a correction to each sim spectra to account for this missing power.  This correction factor is calculated by taking the difference between input CMB theory spectra that have and have not been truncated at $\ell=5100$.  We have verified using Gaussian simulations without missing power, that our {\it{PITAS}} pipeline for computing power spectra is unbiased.  We have also verified that our power spectrum pipeline, using the nominal simulations, is unbiased below $\ell=5000$.  Note that this high-$\ell$ correction factor is only applied to simulated spectra, and not to the data.

We compute the final $TT, TE, EE$, and $BB$ delensed spectra as $C_\ell^{delen} = C_\ell^{raw, delen} - C_\ell^{bias} - C_\ell^{mc}$.  We also calculate the covariance matrix, $\Sigma_{delen}$, from the variance of $C_\ell^{delen}$ using simulation $set_3$.  To this we add the variance of the total bias, $C_\ell^{totbias}=C_\ell^{bias} + C_\ell^{mc}$, calculated as $\Sigma_{totbias} = \Sigma_{bias} + \Sigma_{mc} + 2~Cov(C_\ell^{bias},C_\ell^{mc})$ using simulation $set_1$ and $set_2$.  In addition, we compute the lensed spectra, $C_\ell^{len}$, using the cross spectra of the prepared map splits, and the lensed covariance matrix, $\Sigma_{len}$, from simulations. 
We use these simulation-based $\Sigma_{len}$ and $\Sigma_{delen}$ for the diagonal and $\pm 1$ off-diagonal elements of our final lensed and delensed covariance matrices.

Since the lensing-related off-diagonal components of the covariance matrices from simulations (excluding the $\pm 1$ off-diagonal elements) are not fully converged given our number of simulations, we analytically compute them. In particular, we analytically compute two components, (1)~internal lensing covariance (LC) and (2)~super-sample covariance (SSC). 
The LC term arises because lensing scales within the patch couple together previously-independent CMB modes \cite{Benoitlevy2012,Schmittfull2013,Peloton2017,Green2017}.  For the lensed and delensed spectra, we use the code presented in~\cite{Green2017} to evaluate the LC component using the CMB noise level and lensing map filtering choices from this analysis.  We note that while~\cite{Green2017} found that CMB delensing would essentially remove all internal lens-induced covariance for a futuristic survey, at the noise levels of the current analysis these covariances are reduced by about 50\%.
The SSC term in the covariance matrix originates from CMB modes within the patch coupling to lensing modes larger than our relatively small analysis region, and is therefore not affected by delensing.  The SSC component for both lensed and delensed spectra is analytically computed following the procedure described in~\cite{Manzotti2014,Motloch2019}. We test these analytic LC and SSC components against the off-diagonal covariance matrix terms derived from lensed CMB signal-only simulations, and find a good match.  In particular, we find less than a 5\% difference in off-diagonal correlation matrix elements up to $\ell=3000$; above $\ell=3000$ the lensing induced off-diagonal components become subdominant.  We replace the off-diagonal terms (except the $\pm 1$ off-diagonal elements) in $\Sigma_{delen}$ and $\Sigma_{len}$ with the analytic LC and SSC terms calculated above.

We compute the difference between $C_\ell^{len}$ and $C_\ell^{delen}$ to obtain the difference spectra, $C_\ell^{df}$ = $C_\ell^{len}$ - $C_\ell^{delen}$.  
The covariance matrix of the difference spectra, $\Sigma_{df}$, is computed following the same procedure as $\Sigma_{delen}$ above, except that we do not replace the sim-based off-diagonal terms with the analytic ones.  We expect these off-diagonal terms to converge faster for $\Sigma_{df}$ since there is less scatter from noise and cosmic variance.  We check that these off-diagonals are in fact converged by looking at their behavior as a function of number of simulations.

\subsection{Pipeline Verification}
In order to verify the delensing pipeline, we use simulation $set_3$, which has CMB and $\phi$ realizations that are independent from $set_1$ and $set_2$.  We generate $C_\ell^{delen}$ and $C_\ell^{df}$ for each simulation in $set_3$, using reconstructed $\phi_{\rm{MV}}$ maps from the same simulation set. $C_\ell^{totbias}$ is obtained from simulation $set_1$ and $set_2$, as described above. In Figure~\ref{fig:verification}, we show the mean $C_\ell^{df}$ as the orange points.  Blue triangles show the mean $C_\ell^{df,raw}$ prior to subtracting the biases, $C_\ell^{bias}$ and $C_\ell^{mc}$.  We also separately show the MC bias, $C_\ell^{mc}$, as green stars to give a sense of the size of this correction.{\color{blue}\footnote{Since the MC bias is a small correction, it is only important to include it for the pipeline verification; our error bars for the data are large enough that this MC bias correction is negligible. However, we still correct the data spectra for this MC bias for consistency.}} Here, the black dashed curves show the theoretical expectation for the case of `perfect delensing' assuming no noise; the orange solid curves show the expected delensing achievable given ACT noise levels. These theoretical expectations are calculated using CAMB~\cite{Lewis2000}, given the input theory CMB spectra used to generate the simulations discussed in Section~\ref{sec:sims} and the lensing noise power spectrum discussed in Section~\ref{sec:pipe}.  Using the full covariance matrix, we calculate the probability to exceed the given chi-square (PTE) for the average of the delensed spectra from simulation $set_3$.  These PTEs are calculated with respect to delensed theory curves that are generated with the same cosmology used to make the simulations and the achievable level of delensing given ACT D56 noise levels.  We list these PTE's in Table~\ref{tab:sim}, and find that the average of the delensed spectra are consistent with the theory expectation.

\begin{table}[t]
\vspace{1mm}
\begin{tabularx}{0.346\textwidth}{|c|cc|cc|}
\cline{1-5}
Freq (GHz)& ~Spectra~ & ~PTE~ & ~Spectra~ & ~PTE~ \\
\cline{1-5}
\hline
\hline
$150 \times 150$  & TT & \ptettsimhundredfifty & EE & \pteeesimhundredfifty \\
         & TE & \ptetesimhundredfifty & BB & \ptebbsimhundredfifty \\     
\cline{1-5}
\hline
\hline
$150 \times 98$   & TT & \ptettsimcross & BB & \ptebbsimcross    \\
         & TE & \ptetesimcross & EE & \pteeesimcross  \\    
         & ET & \pteetsimcross & -- & -- \\
\cline{1-5}
\hline
\hline
$98 \times 98$    & TT & \ptettsimninety & EE & \pteeesimninety\\
         & TE & \ptetesimninety & BB & \ptebbsimninety \\
\cline{1-5}
\end{tabularx}
\caption{Shown are the PTEs for the average of the delensed spectra from 512 simulations in $set_3$ described in Section~\ref{sec:sims}.  These PTEs are calculated with respect to delensed theory curves that are generated with the same cosmology used to make the simulations and the achievable level of delensing given ACT D56 noise levels. We find good PTEs for each spectrum type and frequency, indicating the delensing pipeline described in Section~\ref{sec:pipe} is unbiased.}
\vspace{-4mm}
\label{tab:sim}
\end{table}

\section{Delensed Power Spectra}
\label{sec:spec}

We apply the delensing pipeline described in Section~\ref{sec:pipe} to the data described in Section~\ref{sec:data} in order to obtain lensed and delensed spectra, $C_{\ell_j}^{len,data}$ and $C_{\ell_j}^{delen,data}$ where $j \in (TT, TE, EE, BB)$.  We also obtain the difference spectra, $C_{\ell_j}^{df,data}$.

\begin{figure}[t]
  \centering
  \vspace{-1mm}
 \hspace{-2mm}\includegraphics[width=\columnwidth]{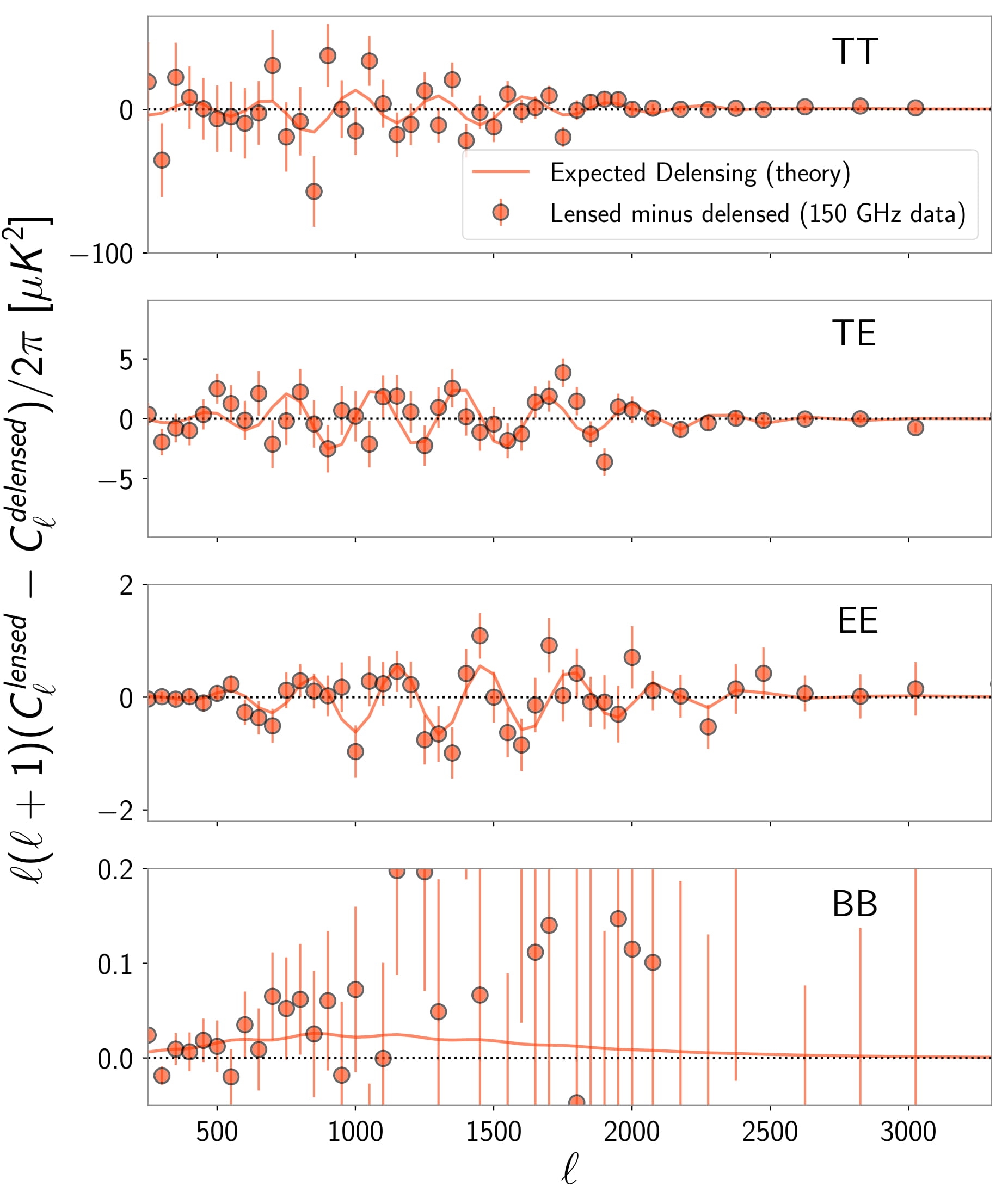}
  \caption{Shown are lensed minus delensed spectra for $150 \times 150$ GHz only, using the data described in Section \ref{sec:data}. The error bars shown are obtained from the diagonal elements of $\Sigma_{df}$, described in Sections~\ref{sec:subBias} and~\ref{sec:spec}. We also show binned theory curves using the parameters obtained from fitting to the lensed and delensed spectra as discussed in Section~\ref{sec:params}. Table~\ref{tab:data} gives the significance with which the delensing effect is detected, using the full covariance matrix, as well as the delensing efficiencies. For $TT$, roughly half the signal-to-noise ratio comes from off-diagonal correlations,  and most comes from roughly $\ell \in [1500,4000]$.  For all the spectra, the modes with $\ell>4000$ contribute very little to the signal-to-noise ratio.}
  \label{fig:data-diff}
\end{figure}

To obtain the final covariance matrices for the data we also add beam uncertainties, calibration uncertainties, and the trispectrum error from Poisson sources largely following~C20. In particular, we calculate the beam uncertainties by first generating a beam realization for each season, array and frequency, from beam profiles described in~A20. These simulated beams are applied to signal-only simulations, which are then processed through our map combining procedure described in Section~\ref{sec:data}. By comparing the beam-convolved output spectra to the original signal-only input spectra, we obtain the effective beam for the combined map.  We repeat this process 1024 times to obtain the mean beam and its error, and we apply the formula in~\cite{Das2014} to calculate the beam covariance matrix, $\Sigma_{beam}$. Similarly, we simulate the mean calibration and its error, and apply the formula mentioned above to obtain the calibration covariance matrix, $\Sigma_{cal}$. The trispectrum component from Poisson sources included in the covariance matrix is identical to that in~C20.

\begin{figure}[t]
  \centering
  \vspace{-1mm}
   \hspace{-3mm}\includegraphics[width=\columnwidth]{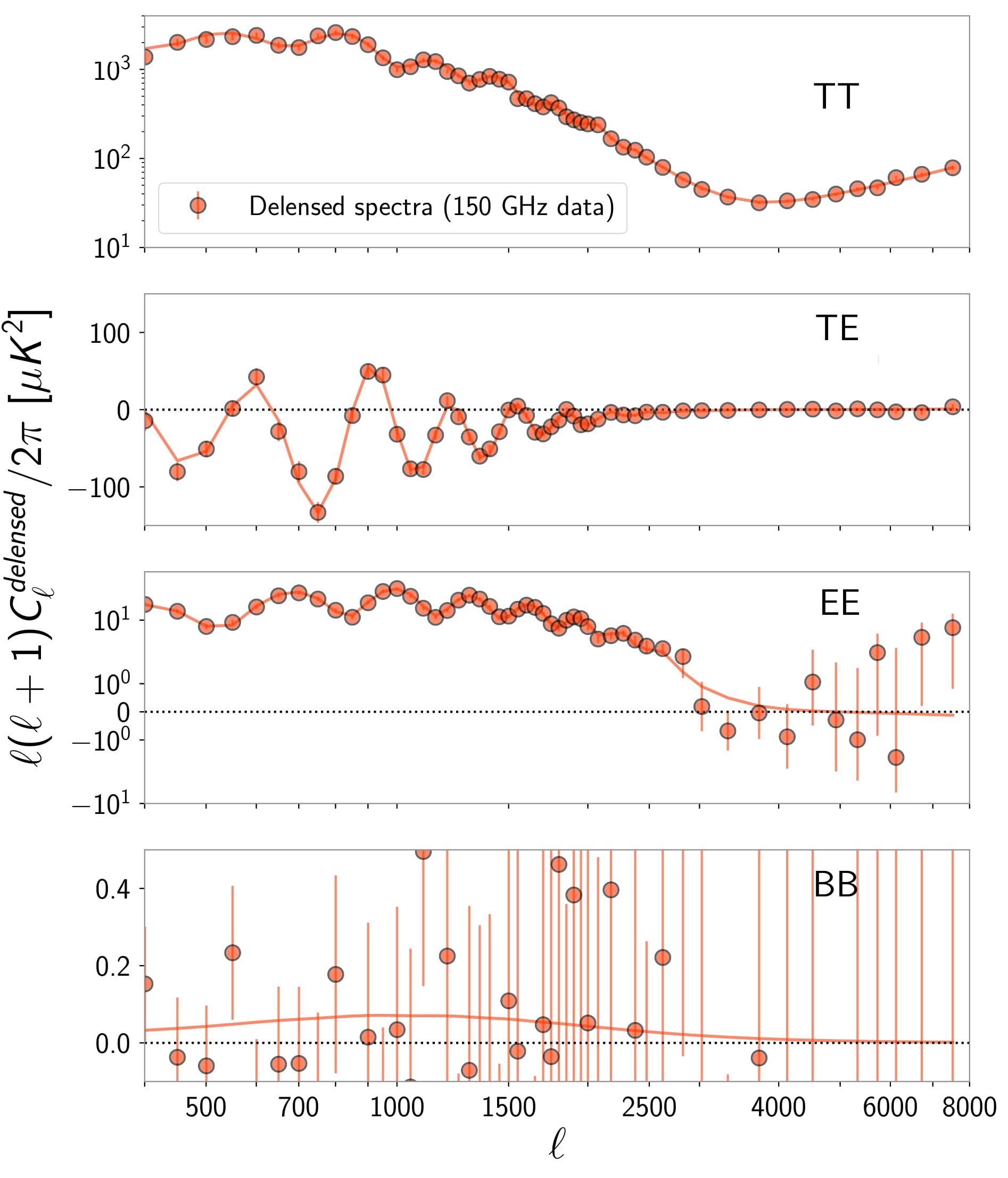}
\caption{Shown are the delensed spectra for $150 \times 150$ GHz only, using the data described in Section~\ref{sec:data}, with error bars from the diagonals of $\Sigma_{delen}$ (described in Sections~\ref{sec:subBias} and~\ref{sec:spec}).  We also show the {\it{binned}} theory curves using the parameters obtained from fitting to the delensed spectra as discussed in Section~\ref{sec:params}.  PTEs with respect to the theory are given in Table~\ref{tab:data}.}
\label{fig:data-delensspec}
\end{figure}

In Figure~\ref{fig:data-diff}, we show the resulting difference spectra of the data, $C_\ell^{df,data}$, for $150 \times 150$ GHz, with the diagonals of $\Sigma_{df}$ as the error bars.  We also obtain lensed and delensed theory spectra, $C_{\ell_j}^{len,th}$ and $C_{\ell_j}^{delen,th}$, from CAMB, as discussed in more detail in Section~\ref{sec:like}, computed assuming the cosmological parameters obtained from fitting to the lensed and delensed spectra as discussed in Section~\ref{sec:params}. The difference of these, $C_\ell^{df, th} = C_\ell^{len,th} - C_\ell^{delen,th}$, is shown as the solid curves in Figure~\ref{fig:data-diff}. We compute the chi-squared with respect to theory for the difference spectra as
\begin{equation}\label{eq:diffchisq}
\chi_{th,df}^2 = (C_\ell^{df, data} - C_\ell^{df, th})^T (\Sigma_{df})^{-1} (C_\ell^{df, data} - C_\ell^{df, th}).  
\end{equation}
We verify with simulation $set_3$ that Eq.~\ref{eq:diffchisq} does follow a $\chi^2$ distribution. The $\chi^2$ for the null-hypothesis, $\chi_{null, df}^2$, is calculated by replacing $C_\ell^{df, th}$ with $C_\ell^{df, null} = 0$ in the equation above. Following~\cite{Hand2015, Madhavacheril2015}, we use a likelihood ratio test to define the significance with which the model fits the data better than a null signal, i.e. the signal-to-noise ratio, S/N, as
\begin{equation}\label{eq:diffsn}
S/N= \sqrt{\chi_{null, df}^2 - \chi_{th, df}^2}. 
\end{equation}
We note that for the detection statistic defined by Eqs.~\ref{eq:diffchisq} and~\ref{eq:diffsn}, much of the common signal and instrument noise in the lensed and delensed power spectra cancels out. This can then yield higher-significance detections of delensing than would be expected without this cancellation. For example, we can detect $BB$ delensing with greater significance than the $BB$ signal itself. This fact has also been used in previous detections of delensing \citep{Larsen2016,Carron2017,Manzotti2017,Planck2018Lensing, Adachi2019}. We find that most of the signal-to-noise ratio is from $\ell$ modes below 4000. For example, for the $150 \times 150$~GHz $TT$ spectrum, the full range of $\ell \in [575, 7925]$ yields a signal-to-noise ratio of $8.7 \sigma$; using the range of $\ell \in [575, 3925]$, yields a signal-to-noise ratio of $7.0 \sigma$.

\begin{table}[t]
\begin{tabularx}{0.455\textwidth}{|c|c|c|c|c|}
\cline{1-5}
Freq (GHz)      & Spectra &~Delens S/N~& Delens Eff.~&PTE \\
\cline{1-5}
\hline
\hline
$150 \times 150$  & TT & $\snrtthundredfifty$ & \efftthundredfifty & \ptettdatahundredfifty \\
         & EE & $\snreehundredfifty$ & \effeehundredfifty & \pteeedatahundredfifty \\ 
         & TE & $\snrtehundredfifty$ & \efftehundredfifty & \ptetedatahundredfifty \\ 
         & BB & $\snrbbhundredfifty$ & \effbbhundredfifty & \ptebbdatahundredfifty \\    
\cline{1-5}
\hline
\hline
$98 \times 150$   & TT & $\snrttcross$  & \effttcross  & \ptettdatacross  \\
         & EE & $\snreecross$  & \effeecross  & \pteeedatacross  \\
         & ET & $\snretcross$  & \effetcross  & \pteetdatacross  \\   
         & TE & $\snrtecross$  & \efftecross  & \ptetedatacross  \\ 
         & BB & $\snrbbcross$  & \effbbcross  & \ptebbdatacross  \\ 
\cline{1-5}
\hline
\hline
$98 \times 98$    & TT & $\snrttninety$ & \effttninety & \ptettdataninety \\
         & EE & $\snreeninety$ & \effeeninety & \pteeedataninety \\
         & TE & $\snrteninety$ & \effteninety & \ptetedataninety \\
         & BB & $\snrbbninety$ & \effbbninety & \ptebbdataninety \\ 
\cline{1-5}
\hline
\hline
All & TT, TE, ET, EE & -- & -- & \pteALL \\
\cline{1-5}
\end{tabularx}
\caption{Shown are the delensing detection significances (calculated from Equation~\ref{eq:diffsn}) and efficiencies ($\epsilon$, calculated using Equation~\ref{eq:eff}) of the data presented in Figure~\ref{fig:data-diff}. Note that we do not detect delensing in the $98 \times 150$~GHz $TE$ spectra. We also list the PTE values of the delensed spectra shown in Figure~\ref{fig:data-delensspec} with respect to theory curves based on the parameters obtained from fitting to the delensed spectra as discussed in Section~\ref{sec:params}. The PTE values are calculated with the covariance matrix $\Sigma_{delen}$ described in Sections~\ref{sec:subBias} and~\ref{sec:spec}, using 47 degrees of freedom for $TT$ and 49 for $TE, EE$, and $BB$.}
\label{tab:data}
\end{table}

\begin{figure*}[t]
\centering
  \includegraphics[width=1.0\textwidth]{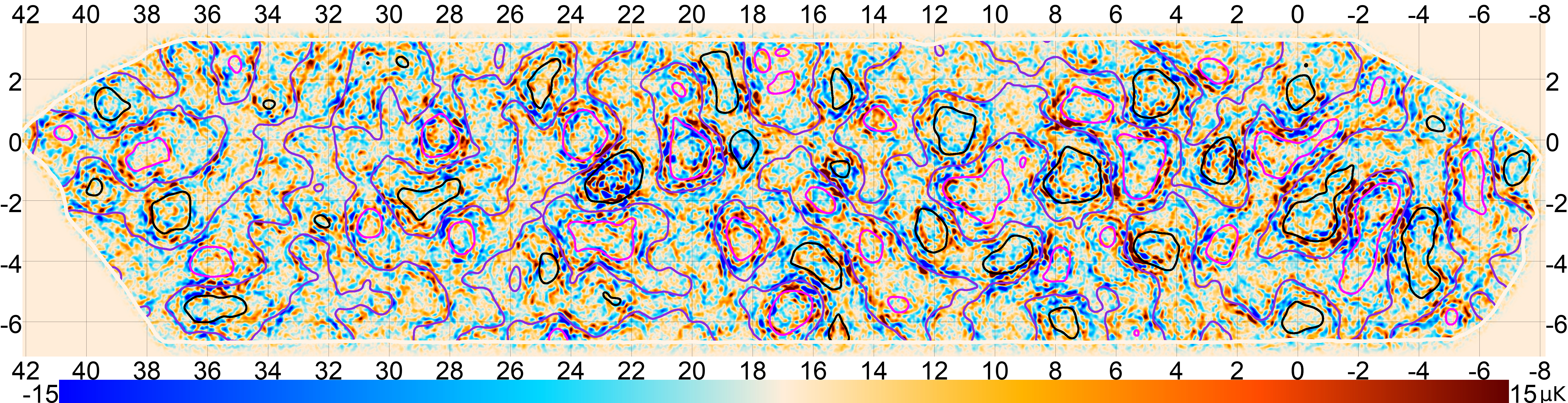}
  \caption{Shown is an image of the lensed minus delensed CMB map from the ACT D56 150 GHz temperature data.  Here we coadd two splits of the lensed and delensed data before differencing. We also Wiener filter the difference map to downweight the noisy modes. This image shows the familiar distortions caused by gravitational lensing of the CMB. In addition, we overlay the contours of the reconstructed minimum variance ACT D56 lensing potential map ($\phi_{\rm{MV}}$) obtained as described in Sections~\ref{sec:reconstruction} and~\ref{sec:delensing}. The black contours indicate regions of overdensity, while the magenta contours show regions of underdensity in the potential map. As shown, the regions with the steepest contour gradients line up with the largest residuals in the CMB difference map.} 
  \label{fig:LensedMinusDelensed}
\end{figure*}

Following Eq.~3.6 in~\cite{Carron2017}, we also calculate the delensing efficiency, $\epsilon$, for each spectrum by minimizing
\begin{equation}\label{eq:eff}
\chi^2(\epsilon) = \sum_{l} [C_\ell^{df, data} - \epsilon(C_\ell^{len,th}-C_\ell^{unlen,th})]^2/\sigma^2_l 
\end{equation}
for $\ell \in [575, 3125]$ for $TT$, and $\ell \in [475, 3125]$ for $TE$ and $EE${\color{blue}\footnote{We set $\ell_ {max}=3125$ here because technically $\epsilon$ has an $\ell$-dependence that becomes more pronounced as we include higher~$\ell$.}}, where $\sigma_l^2$ are the diagonal elements of $\Sigma_{df}$.  To calculate $C_\ell^{unlen,th}$, we generate unlensed spectra with CAMB, using the parameters obtained from fitting to the lensed spectra as discussed in Section~\ref{sec:params}. 
The resulting delensing signal-to-noise ratios and delensing efficiencies ($\epsilon$) for the data are listed in Table~\ref{tab:data}.

In Figure~\ref{fig:data-delensspec}, we show $C_\ell^{delen,data}$ for the data for $TT, TE, EE,$ and $BB$ at $150 \times 150$ GHz, plotting the diagonals of $\Sigma_{delen}$ as the error bars. We show in Table~\ref{tab:data} the PTEs for all the individual delensed spectra compared to theory curves derived from the parameters obtained from fitting the delensed spectra as discussed in Section~\ref{sec:params}; the PTEs are computed assuming 47 degrees of freedom for $TT$ and 49 for $TE, EE$, and $BB$, the number of spectral bins in each case. In Sections~\ref{sec:like} and~\ref{sec:params}, we discuss in detail how we use these delensed spectra to obtain cosmological parameters.  For obtaining the total PTE from all the spectra (``All'' in Table~\ref{tab:data}), we use a total of $484-20=464$ degrees of freedom (484 spectral bins minus 20 free parameters) when fitting all the spectra together.  Note that we do not detect delensing in the $98 \times 150$~GHz $TE$ spectra, and the delensing signal-to-noise ratio and the delensing efficiency are only calculated for individual spectra.

In Figure~\ref{fig:LensedMinusDelensed}, we show for purely visualization purposes an image of the lensed CMB minus the delensed CMB from our data.  Here we coadd the two splits of the lensed and delensed data before differencing. We also Wiener filter the difference map to downweight the noisy modes. In this image, we see the familiar distortions caused by gravitational lensing.{\color{blue}\footnote{We note that since the lensing potential map has large-scale lensing modes of $L<80$ removed, we do not see the largest-scale filamentary structures in this CMB difference map.}}  We also overlay the contours of the reconstructed minimum variance lensing potential map ($\phi_{\rm{MV}}$) obtained as described in Sections~\ref{sec:reconstruction} and~\ref{sec:delensing}. The black contours indicate regions of overdensity and the magenta contours regions of underdensity in the potential map. We see that the regions with the steepest contour gradients in the $\phi_{\rm{MV}}$ map line up with the largest residuals in the lensed minus delensed CMB map.

\section{Likelihood for Delensed Spectra}
\label{sec:like}
We use the delensed spectra, $C_{\ell_j}^{delen, data}$, from Section~\ref{sec:spec} for $j \in (TT, TE, EE)$ in the likelihood analysis to obtain cosmological parameters.  To generate appropriate delensed theory spectra, $C_{\ell_j}^{delen, th}$, given our ACT lensing noise levels, for a given parameter set from CAMB, we follow the procedure in~\cite{Carron2017}. Specifically, we first obtain $C^{\kappa\kappa}(L)^{th}$ and $C_\ell^{unlen,th}$ for that parameter set. We then Wiener filter $C^{\kappa\kappa}(L)^{th}$ using the lensing noise $N(L)_{MV}$ described in Section~\ref{sec:pipe}, and calculate the residual lensing power:
\begin{equation}
{C}^{\kappa\kappa}(L)^{th, res} = {C}^{\kappa\kappa}(L)^{th}~\left[1-\frac{C^{\kappa\kappa}(L)^{th}}{C^{\kappa\kappa}(L)^{th} + N(L)_{MV}}\right]. 
\end{equation}
Inputting ${C}^{\kappa\kappa}(L)^{th, res}$ and $C_\ell^{unlen,th}$ into CAMB yields $C_\ell^{delen,th}$. We cross check our ``CAMB-derived'' $C_\ell^{delen,th}$ against similar code used in~\cite{Green2017}, and find excellent agreement. Assuming Gaussian uncertainties on the delensed spectra and using $\Sigma_{delen}$ from Section~\ref{sec:subBias}, we write the log-likelihood as: 
\begin{equation}
\begin{split}
& -2{\rm{ln}} \mathcal{L}  =  \\
& (C_{\ell_j}^{delen,data}-C_{\ell_j}^{delen,th})^T\Sigma_{{delen}_{j,j'}}^{-1}(C_{\ell_{j'}}^{delen,data}-C_{\ell_{j'}}^{delen,th}).
\end{split}
\end{equation}

\subsection{Cosmology Dependence of $C_\ell^{bias}$}
\label{sec:Clbias}
We compute the delensing bias, $C_\ell^{bias}$, with simulations that are based on the {\it{Planck}} cosmology.  To make sure the likelihood given above is sufficiently accurate, we investigate the dependence of $C_\ell^{bias}$ on cosmological parameters.  
We do this by generating flat-sky periodic simulations for a D56-sized sky patch at 150 GHz following a more simplified procedure than described in Section~\ref{sec:sims}; namely, the simulations include $10~\mu$K-arcmin white noise, are convolved with a 1.3 arcminute Gaussian beam, and do not contain any foregrounds. We generate these simulations at several additional cosmologies: (i)~one with the parameters set to the {\it Planck} values from {\it{TT+lowP}} (obtained using the Plik likelihood and varying 6 $\Lambda$CDM parameters)~\cite{Planck2015Parameters}, and (ii)~ten with parameters drawn randomly from the bottom 30\% of a converged {\it Planck} 2015 parameter chain (also generated using {\it{TT+lowP}} and varying 6 $\Lambda$CDM parameters).  We find that the difference in $C_\ell^{bias}$ between case (i) and the ten cases of (ii), is generally less than $5\%$ of the relevant error bars for all bandpowers in the $\ell$ range used in this cosmological analysis. (The error bars being compared to are the square root of the diagonal elements of $\Sigma_{delen}$.) For our most deviant cosmology case, the deviations away from the fiducial $C_\ell^{bias}$ are less than $20\%$ of the relevant error bars, and the sum of the fractional deviations, added in quadrature over the spectral bins, is less than $40\%$. 

We also check that the delensing bias does not have significant cosmology dependence for beyond-$\Lambda$CDM cosmologies. We repeat the test described above for two extended $\Lambda$CDM cosmologies that also vary 1) the effective number of relativistic species and the running of the scalar spectral index, or 2) the effective number of relativistic species and the sum of neutrino masses. We use the {\it Planck} 2018 {\it TT,TE,EE+lowE}~\cite{Planck2018Parameters} best-fit parameters to generate simulations and obtain the delensing bias for each extended cosmology. For consistency with the analysis above, we compare each to the delensing bias obtained from simulations generated using the {\it Planck} 2015 {\it TT+lowP} $\Lambda$CDM parameters~\cite{Planck2015Parameters}. For these extended cosmologies, we find deviations in the delensing bias consistent with those found in the test described above.

In addition, we take simulations that have {\it{WMAP5}} cosmology~\cite{Komatsu2009}, and perform the same delensing procedure, except that we use our original $C_\ell^{bias}$ computed from {\it{Planck2015}}-based simulations.{\color{blue}\footnote{The {\it{WMAP5}}-based simulations use $\Omega_b h^2=0.02218$, $\Omega_c h^2 = 0.1109$, $h=0.71$, $\tau = 0.087$, $A_s=2.45\times 10^{-9}$, and $n_s=0.96$.  Here we take $k_0 = 0.002$ Mpc$^{-1}$ as the pivot scale and the total mass of neutrinos as 0.0 eV.}}  If $C_\ell^{bias}$ has a noticeable cosmology dependence, we would obtain incorrect delensed spectra that do not match the theory expectation.  However, when we compute the delensed spectra with these {\it{WMAP5}} simulations and calculate the PTE with respect to the best-fit delensed theory curve, we find good agreement with a PTE value of 0.29.
Thus we neglect the cosmology dependence of $C_\ell^{bias}$.

\begin{figure}[t]
\centering
  \vspace{-1mm}
  \includegraphics[width=0.47\textwidth]{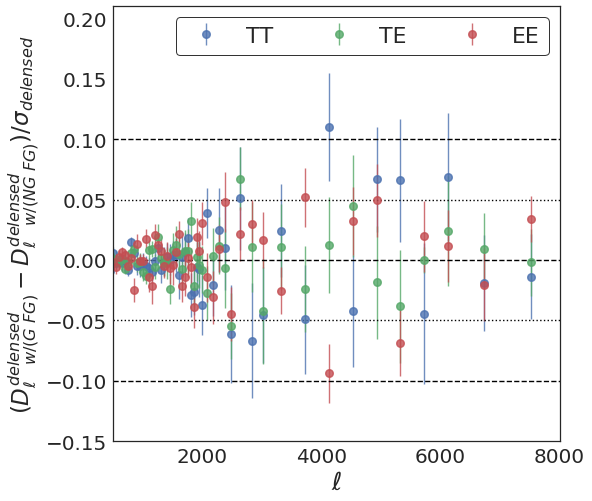}
  \caption{Difference in the resulting delensed spectra with Gaussian and non-Gaussian foregrounds applied to the same CMB realization. We show the mean difference and the scatter of 72 simulations discussed in Section~\ref{sec:FG} as a fraction of the error bar on the delensed D56 data spectra. Each simulation has an independent CMB realization with separate Gaussian and non-Gaussian foreground realizations and a size of about 100 square degrees. The horizontal dotted and dashed black lines indicate a difference that is 5\% and 10\%, respectively, of the error bar. We find that the difference is less than $10\%$ of the error bar on the delensed D56 data spectra with no evidence of bias.}  
  \label{fig:non-G-FG}
\end{figure}

\begin{figure*}[t]
  \centering
  \includegraphics[width=1.0\textwidth]{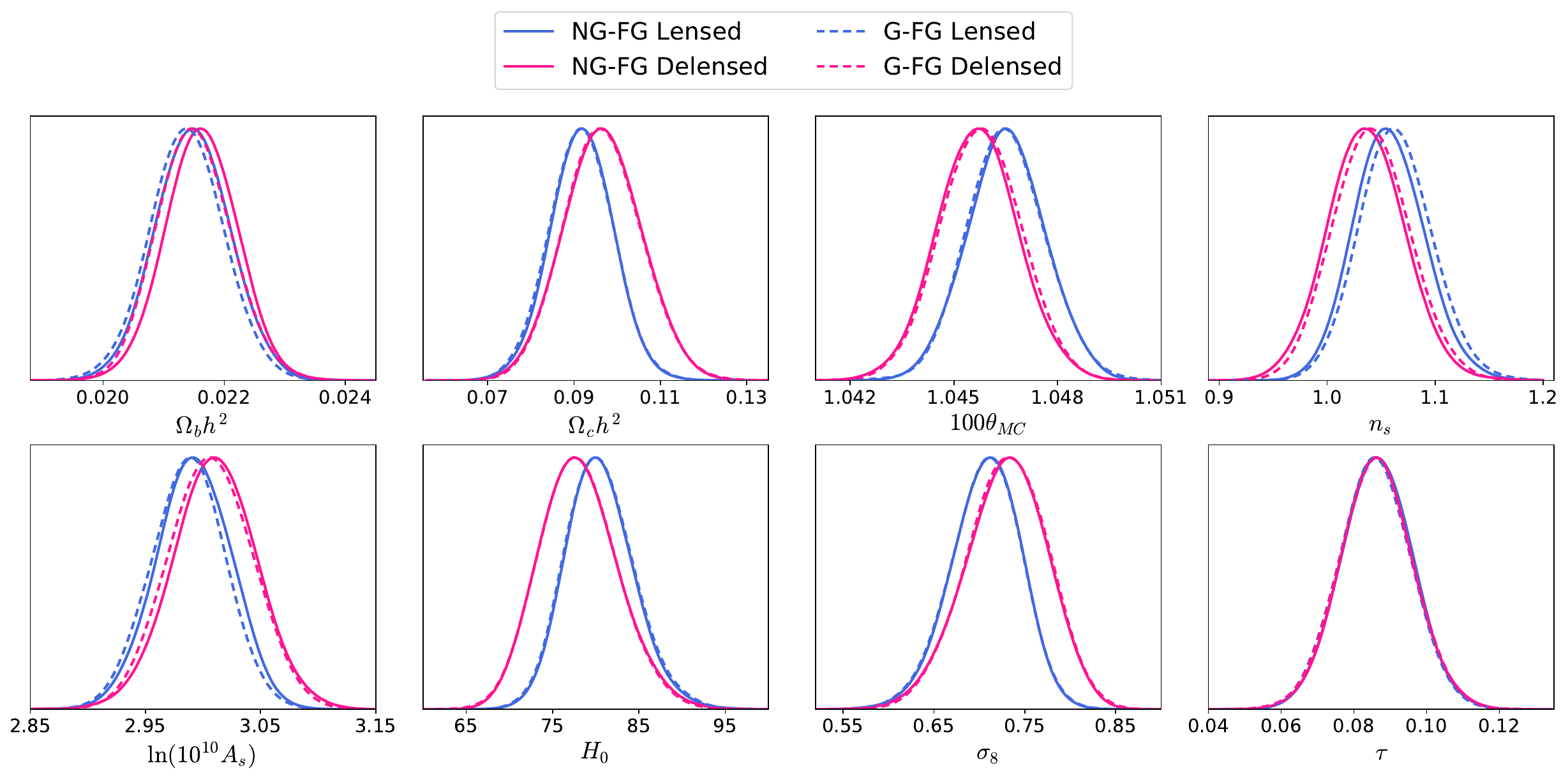}
  \caption{Here we test the impact of non-Gaussian foregrounds on our delensing analysis using a D56-sized simulation from~\cite{Sehgal2010}, described in Section~\ref{sec:FG}. Shown are parameter constraints from the spectra of these simulations, with both Gaussian (G, dashed curves) and non-Gaussian (NG, solid curves) foregrounds applied to the same CMB realization.  This plot shows that the parameter differences between delensed spectra using either G or NG foregrounds (e.g.~the difference between the dashed and solid curves of the same color) are typically smaller than the parameter differences between lensed and delensed spectra (e.g.~the difference between the magenta and blue curves).}
  \label{fig:NGvsGFGparams}
\end{figure*}

\subsection{Impact of Foregrounds}
\label{sec:FG}
Another potential source of systematic effects is the impact of foregrounds.  For example, it is possible that (i)~the foreground spectra may change appreciably during the process of delensing.  Another potential systematic effect is that (ii)~the delensed spectra might be biased because we used reconstructed potential maps that have residual foregrounds in them that are correlated with the foregrounds in the maps we are delensing, and with the lensing potential itself. A final possible systematic is that (iii)~the non-Gaussian nature of the foregrounds themselves could add bias to the lensing reconstruction and delensed spectra.

To test these potential systematic effects, we use simulations from~\cite{Sehgal2010} that include non-Gaussian distributions of extragalactic foregrounds that are correlated with the lensing potential. The amplitudes of the non-Gaussian foregrounds in these simulations have been adjusted to match recent CMB data; these adjusted simulations were used in the Simons Observatory (SO) forecasting paper~\cite{SO2019} and are public at \url{https://lambda.gsfc.nasa.gov/toolbox/tb_cmbsim_ov.cfm}.

We start by regenerating new unlensed CMB maps{\color{blue}\footnote{We regenerate new CMB maps since the original simulations do not include lensed CMB polarization maps.}} with the input CMB power spectra from~\cite{Sehgal2010}, and then lens them with the kappa map provided in~\cite{Sehgal2010}.  We then add to each CMB realization a noise realization matching our data. To each CMB plus noise realization, we then add (1)~a realization of non-Gaussian (NG) foregrounds from~\cite{Sehgal2010}, or (2)~a realization of Gaussian (G) foregrounds, both with identical foreground flux cuts of 15 mJy at 150 GHz and identical total foreground power spectra (made to match by construction). We run these CMB plus noise realizations with G and NG foregrounds through the delensing pipeline described above.  This is repeated for 72 independent CMB plus noise realizations with a footprint-size of about 100 square degrees. Each of the 72 independent CMB realizations has a separate G and NG foreground realization.  A bias would show up as a difference in delensed spectra with G versus NG foregrounds, the latter of which have residual foregrounds correlated with the lensing potential (systematic ii), non-Gaussian structure (systematic iii), and may have a different behavior when undergoing the process of delensing than G foregrounds (systematic i).

To quantify the difference in delensed spectra for the case with G versus NG foregrounds, we first explicitly correct both the lensed and delensed spectra for the difference between G and NG foreground power spectra from each lensed realization.  Specifically, we add $\Delta C_\ell^{FG}$ = $C_\ell^{len,G}-C_\ell^{len,NG}$ to $C_\ell^{len,NG}$ and $C_\ell^{delen,NG}$. Therefore, at the power spectrum level, the {\it{lensed}} spectra for each CMB realization with either G or NG foregrounds matches by construction.  
We then difference the resulting delensed spectra with G and NG foregrounds, and find that the difference is less than 5 to 10$\%$ of the error bars on the delensed spectra, with no evidence of bias, as shown in Figure~\ref{fig:non-G-FG}. 

Additionally, as a more stringent test, we check the resulting parameters obtained from the delensed G and NG power spectra, this time using D56-sized patches, as shown in Figure~\ref{fig:NGvsGFGparams}. In this case, we also apply a 5~mJy flux cut for both G and NG foregrounds in the simulations used to make the lensing reconstruction (to match the $5\sigma$ flux cut we apply to the data).  We also do not force each realization of the lensed G and NG foreground spectra to match by construction.  Figure~\ref{fig:NGvsGFGparams} shows the parameter results from one of our two D56 simulated patches, where we have either G or NG foregrounds when constructing the delensed spectra.  We see that the differences in marginalized mean parameters obtained from the delensed spectra using either G or NG foregrounds is in general smaller than the parameter differences between lensed and delensed spectra.  We also find the same to be true for the best-fit parameters discussed further in Section~\ref{sec:param-shift-covmat}.  Thus we neglect any bias from using G as opposed to NG foregrounds in our simulations, and more broadly we neglect any bias related to delensing foregrounds.

\vspace{-2mm}
\section{Cosmological Parameters}
\label{sec:params}
The Monte Carlo sampler CosmoMC~\cite{Lewis2002} is used to find the marginalized mean cosmological parameters from the likelihood for the delensed spectra discussed in Section~\ref{sec:like}.  We vary the $\Lambda$CDM parameters $\Omega_b h^2$, $\Omega_c h^2$, $\theta_{MC}$, $\ln (10^{10} A_s)$, $n_s$, and $\tau$, and adopt the same 14 foreground and calibration parameters as in~C20 and~A20.  We use the same prior ranges on these parameters as in~C20 and~A20, which we list in Table~\ref{tab:priors}.

We also set the effective number of relativistic species, ${N}_{\rm{eff}}$, equal to 3.046, the dark energy equation state, $w$ equal to $-1$, the sum of the neutrino masses $\Sigma m_{\nu}$ equal to 0.06 eV, and the pivot scale $k_0$ equal to $0.05 \ \textrm{Mpc}^{-1}$. The helium fraction, $Y_p$, is set assuming BBN consistency within CosmoMC.   We also find marginalized mean parameters for the lensed spectra, using the same priors as mentioned above.  

\begin{table}[t]
\begin{centering}
\begin{tabularx}{0.46\textwidth}{|cc|cc|}
\cline{1-4}
~~Parameter    ~~       & Prior    ~~~     & Parameter           & Prior          \\
\cline{1-4}
~~$\Omega_b h^2$  ~~   & [0.005, 0.1]  ~~~ & $\ln(10^{10} A_s)$ & [2, 4]  \\
~~$\Omega_c h^2$  ~~   & [0.001, 0.99]  ~~~   &  $n_s$ & [0.8, 1.2]  \\ 
~~$100\theta_{MC}$ ~~  &  [0.5, 10.0]   ~~~  & $\tau$ &  $0.065 \pm 0.015$ ~~ \\
~~$A_{s,d}$  ~~   & $3.1 \pm 0.4$ ~~~ & $A_c$ & $4.9 \pm 0.9$  \\
~~$A^{TE}_{PS}$  ~~   & [-1, 1]  ~~~   &  $\beta_c$           & [0, 5]  \\ 
~~$A^{EE}_{PS}$ ~~  &  [0, 1]   ~~~  & $A_d$     & [0, 11] \\
~~$A^{TT}_{dust,d}$  ~~   & $2.79 \pm 0.45$  ~~~ &~~~  ~~$A_{tSZ}, A_{kSZ}$ & [0, 10] \\
~~$A^{TE}_{dust,d}$  ~~   & $0.11 \pm 0.10$ ~~~ &  $\xi$   & [0, 0.2] \\
\rule[-1.2ex]{0pt}{0pt} ~~$A^{EE}_{dust,d}$  ~~   &  $0.04 \pm 0.08$ ~~~ & $y^{P}_{98},y^{P}_{150}$ & [0.9, 1.1] \\ 
\cline{1-4}
\end{tabularx}
\caption{Shown are the prior ranges for the 20 parameters used in the parameter analysis. Priors in brackets are flat priors, and the others are Gaussian priors with the given center and $1\sigma$ uncertainty. Priors on all the parameters match those in~C20, which also defines the 14 parameters related to foregrounds and calibration.}
\label{tab:priors}
\end{centering}
\end{table}

\subsection{Parameter-shift Covariance Matrix} \label{sec:param-shift-covmat}

Since we separately obtain parameters for delensed spectra and lensed spectra, we can also measure the shift between these parameters.  When fitting with the correct model, we expect the ensemble average of the parameter shifts to be zero.  In addition, since both the lensed and delensed spectra are sourced from the same region of sky, uncertainty in this shift is minimized due to sample variance cancellation.  As mentioned in Section~\ref{sec:intro}, a significant non-zero shift in parameters would suggest a failure of the model being fit, and could indicate either new physics at early times or a systematic effect.

To assess whether any parameter shift we find from the data is as expected, we need to obtain a covariance matrix for the parameter shifts using lensed and delensed spectra from simulations.  Since obtaining a parameter-shift covariance matrix from hundreds of CosmoMC parameter chain runs would be computationally infeasible, we instead use the $action=2$ setting in CosmoMC to determine global best-fit parameter values.  This $action=2$ setting allows us to start at four random initial positions in parameter space, and iteratively maximize the likelihood, checking that the four initial positions converge to the same maximum likelihood point.
We find that the difference between the $\Lambda$CDM marginalized mean parameters from a full CosmoMC chain ($action=0$) and the best-fit parameters ($action=2$) is less than $2\%$ for each parameter, except for $\tau$ (the least constrained parameter) which can differ by up to $6\%$. This random scatter between $action=2$ and $action=0$ parameters can be reduced further by averaging a number of $action=2$ runs together for the same spectra.  We show in Figure~\ref{fig:action=2Convergence} in Appendix~\ref{sec:appendixParamShiftCov} that averaging twenty $action=2$ runs together for a given spectra is sufficient to achieve convergence in cosmological parameters.  When we do this, the difference between $\Lambda$CDM parameters from $action=2$ versus $action=0$ is often less than $0.5\%$. (We note that in general marginalized means and best-fits should only agree exactly when the distributions are multi-variate Gaussian.)  Thus, in this analysis, the best-fit parameters for a given spectra set are obtained from an average over twenty $action=2$ runs.  Similarly, the parameter-shift covariance matrix (discussed below) is also computed by averaging twenty $action=2$ runs for each of the lensed and corresponding delensed spectra from simulation $set_3$.  

\begin{figure}[t]
  \centering
\hspace{-4mm}  \includegraphics[width=0.5\textwidth]{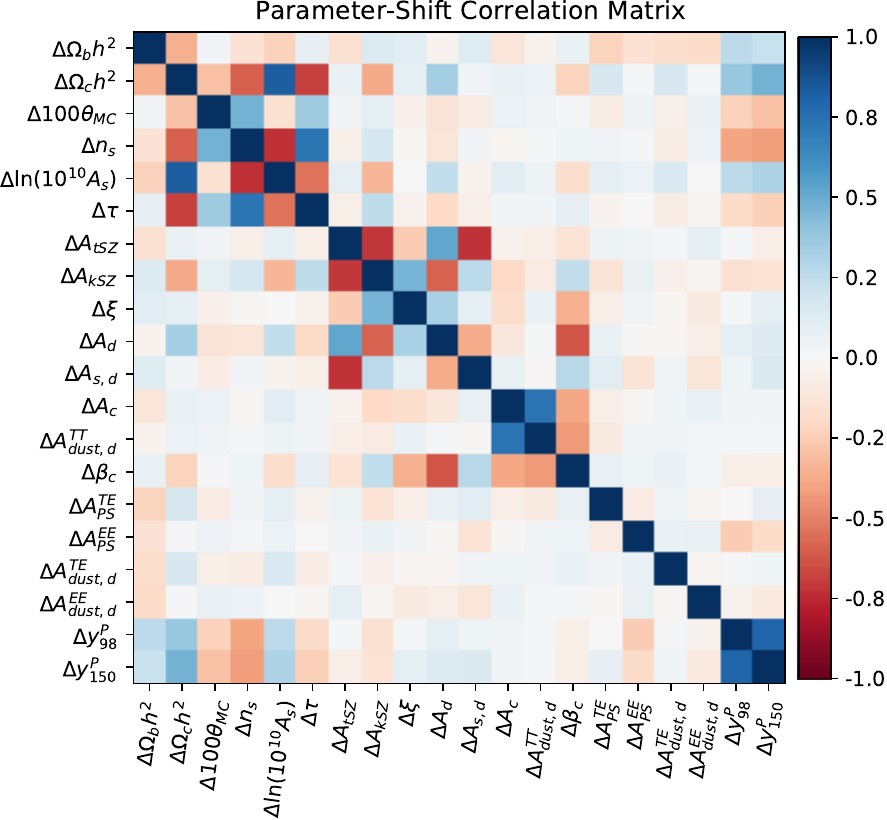}
  \caption{Shown is the full final 20x20 parameter-shift covariance matrix, described in Section~\ref{sec:param-shift-covmat} and used throughout this work, represented as a correlation matrix to make the structure more apparent. This covariance matrix includes the 20 parameters (6 $\Lambda$CDM parameters and 14 foreground and calibration parameters defined in C20) varied in the parameter analysis described in Section~\ref{sec:params}. Each element of the covariance matrix is the variance/covariance of the difference in inferred best-fit parameters derived from lensed and delensed spectra obtained from 300 simulations.}
  \label{fig:full-param-shift-covmat}
\end{figure}

To compute the parameter-shift covariance matrix, we first compute the difference in best-fit parameters between lensed and delensed spectra from each of 300 simulation realizations in $set_3$.  We denote this shift within the space of the 20 parameters as $\vec \theta^\mathrm{shift}$. From these differences, we obtain the parameter-shift covariance matrix, $\mathbb{C}^\mathrm{shift}$, according to 
\begin{equation}
    \mathbb{C}_{ab}^\mathrm{shift} = \langle \vec \theta_a^\mathrm{shift} \vec \theta_b^\mathrm{shift} \rangle, 
    \label{eq:paramshiftcovmat}
\end{equation}
where the angled brackets denote the average over the 300 simulations.   This covariance matrix, shown in Figure~\ref{fig:full-param-shift-covmat}, includes the 14 foreground and calibration parameters (bottom right blocks) in addition to the 6 $\Lambda$CDM parameters (top left block).

Given the shift in the best-fit parameters from delensing obtained with the data, $ \vec {\theta}^\mathrm{shift,data}$, we then define a $\chi^2$ statistic given by 
\begin{equation}
\chi^2 = ({ \vec {\theta}^\mathrm{shift,data}})^T \left (\mathbb{C}^\mathrm{shift}\right)^{-1} \vec \theta^\mathrm{shift,data} 
\label{eq:chi2paramshift}
\end{equation} which we will use below to assess whether there are appreciable shifts in parameters between the lensed and delensed datasets.   We note that for this definition, our theory expectation is that there is no appreciable shift after delensing is applied, i.e. $\vec \theta^\mathrm{shift,theory} = 0$.{\color{blue}\footnote{In principle, various lensing approximations, such as using anti-lensing instead of inverse-lensing (see footnote 8), might induce a non-zero, albeit very small, shift in parameters. However, these secondary shifts should be corrected for by the MC bias we discuss in Section~\ref{sec:subBias}.}} We show explicitly with simulations in Figures~\ref{fig:ParamShiftChi2} and~\ref{fig:AL} in Appendix~\ref{sec:appendixParamShiftCov} that Eq.~\ref{eq:chi2paramshift} follows a $\chi^2$ distribution, and in Figure~\ref{fig:ParamShiftBias2} that there is no significant bias away from the expectation of a mean shift of zero.
\\

{\it{Dependence on Cosmology:}} To check for any dependence on cosmology of the parameter-shift covariance matrix, we perform a test where we use lensed and delensed spectra from a CMB simulation that uses the {\it{WMAP}} 5-year best-fit cosmology~\cite{Komatsu2009}, instead of the {\it{Planck}} best-fit cosmology on which the covariance matrix is based.  We then run the {\it{WMAP5}}-based lensed and delensed spectra through the cosmology analysis, and find that the shifts between lensed and delensed parameters are consistent with the expectations of the covariance matrix, yielding a PTE of $\WMAPpteparamshift$ (20 dof).  Hence we conclude that there is negligible cosmology dependence of our parameter-shift covariance matrix.  \\

{\it{Addition of Galactic Dust:}} Since we generate the parameter-shift covariance matrix using simulations with no Galactic dust, we also check whether the inclusion of Galactic dust would generate a failing PTE for this parameter-shift statistic.  We add to one simulation realization from $set_3$ a realization of Galactic dust for temperature maps obtained from the simulation of~\cite{Sehgal2010}.  Running this simulation including Galactic dust through our delensing pipeline, we find shifts between lensed and delensed parameters consistent with our parameter-shift covariance matrix, yielding an acceptable PTE of $\GalDustpteparamshift$ (20 dof).  Thus we find that the omission of non-zero Galactic dust foregrounds in the simulations used to generate the parameter-shift covariance matrix does not result in a failure of the parameter-shift test when applied to data that includes Galactic dust foregrounds. \\

\begin{figure*}[t]
  \centering
\includegraphics[width=1\textwidth]{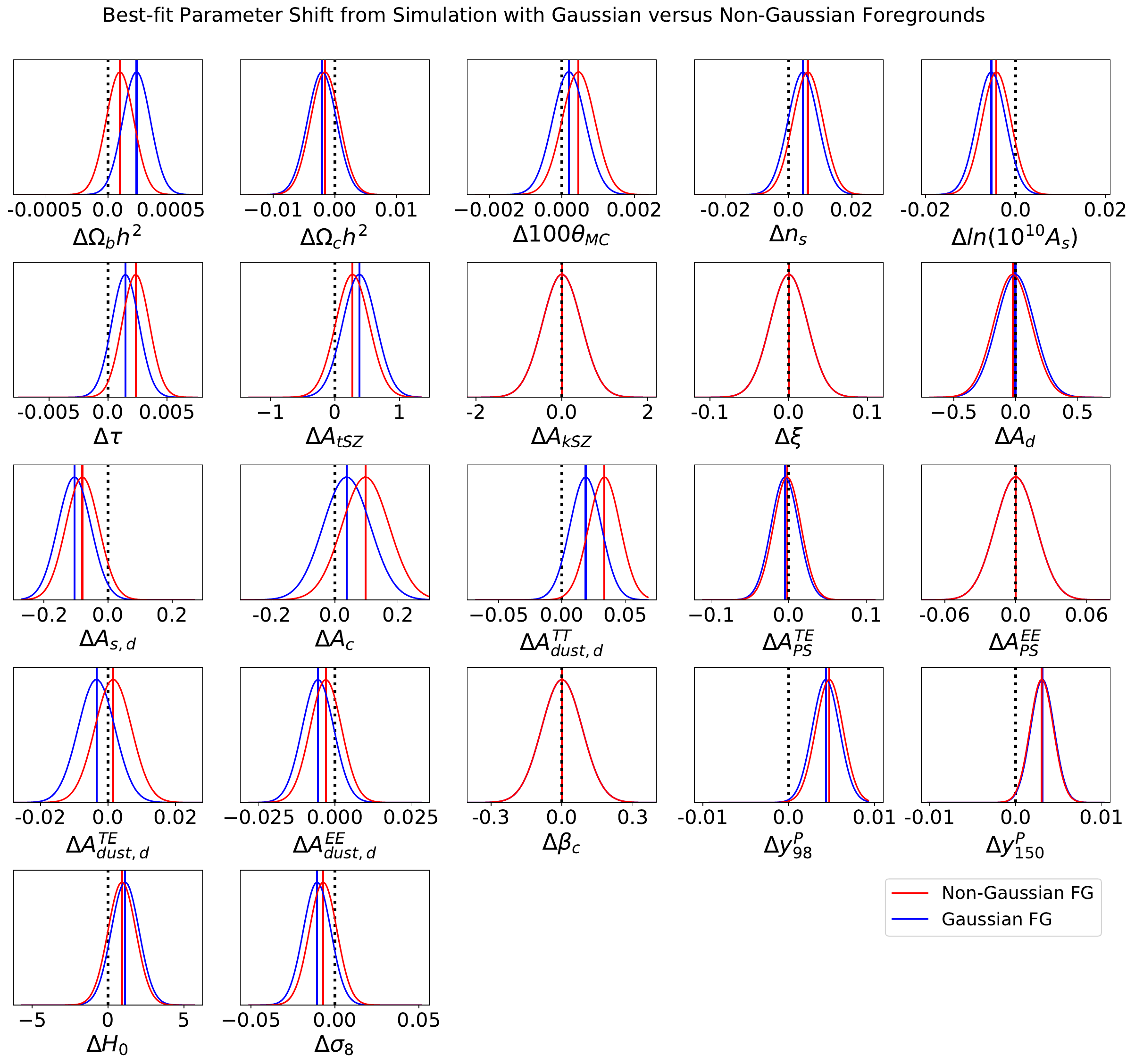}
  \caption{Shown are the lensed minus delensed shifts in best-fit parameters derived from spectra obtained from the CMB plus foreground simulation described in~\cite{Sehgal2010} and in Section~\ref{sec:FG}. The red vertical lines show the shifts from these simulations that include non-Gaussian foregrounds that are also correlated with the lensing potential.  The blue vertical lines show the parameter shifts when the foreground part of the simulation is replaced with Gaussian foregrounds that have similar (but not exact) power spectra as the non-Gaussian foregrounds. Since using twenty $action=2$ runs does not yield the full posterior distributions, the red/blue curves show Gaussian distributions centered on the red/blue vertical lines with widths determined by the diagonal elements of the parameter-shift covariance matrix, as described in Section~\ref{sec:param-shift-covmat}. The black dotted line indicates zero shift in parameters. Even though the foregrounds in this simulation are not exactly matched to the foreground model used to generate the parameter-shift covariance matrix, both Gaussian and non-Gaussian models have acceptable PTEs: $\Gpteparamshift$~($\Gsnrparamshift$)~and $\NGpteparamshift$~($\NGsnrparamshift$)~for Gaussian and non-Gaussian models respectively. Moreover, the difference between Gaussian and non-Gaussian shifts is small, suggesting that the non-Gaussian nature of the true foregrounds does not adversely affect this parameter-shift statistic.}
  \label{fig:paramShiftFG}
\end{figure*}

{\it{Impact of Non-Gaussian Foregrounds:}} In addition, we show in Figure~\ref{fig:paramShiftFG} the sensitivity of the parameter-shift statistic to a non-Gaussian versus Gaussian foreground model.  We do this by obtaining shifts in best-fit parameters from lensed versus delensed spectra obtained from the CMB plus foreground simulation of~\cite{Sehgal2010}, described in Section~\ref{sec:FG}. The red curves show the shifts from these simulations that include non-Gaussian foregrounds that are also correlated with the lensing potential.  We then replace the foreground part of this simulation with Gaussian foregrounds that have power spectra matched to the non-Gaussian foregrounds, when both spectra are calculated over the full-sky (however, they do not identically match in the D56-sized patch used here).  The blue curves show the parameter shifts when using this Gaussian foreground model.  Even though the foregrounds in this simulation are not exactly matched to the foreground model used to generate the parameter-shift covariance matrix, both Gaussian and non-Gaussian models have acceptable PTEs, which are  $\Gpteparamshift$~($\Gsnrparamshift$ for 20 dof)~and $\NGpteparamshift$~($\NGsnrparamshift$ for 20 dof)~for Gaussian and non-Gaussian models respectively. More importantly, there is little difference between the Gaussian and non-Gaussian shifts shown in Figure~\ref{fig:paramShiftFG}, indicating that non-Gaussian correlated foregrounds do not adversely impact this parameter-shift statistic.

\subsection{$\Lambda$CDM Lensing Consistency Test}
\label{sec:newPhysics}

To test our sensitivity to physics outside of the $\Lambda $CDM model, we implement a toy model in which there is an anomalously high lensing-like signal in the CMB power spectrum that is not consistent with the lensing power spectrum within the framework of $\Lambda$CDM.  This demonstration model is in part motivated by the recent {\it Planck} lensing results that found excess smoothing in their CMB power spectra~\cite{Planck2018Parameters}. Another motivation is that several models of interest that differ from $\Lambda $CDM include a feature that can mimic lensing-induced peak smoothing, as discussed in~\cite{Hazra2014,Munoz2016,Smith2017,Hazra2019,Knox2019,Domenech2019,Domenech2020}. 

To implement this toy model we obtain best-fit parameter shifts from lensed versus delensed spectra, using a simulation from $set_3$.  We then add extra smoothing that matches $20\%$ of the lensing-induced peak smoothing (i.e.,~we make $A_L=1.2$~\cite{Calabrese2008}), by adding $0.20\times (C_\ell^{len,th}-C_\ell^{unlen,th})$ to both lensed and delensed spectra.  We find that this extra smoothing shifts the cosmological parameters in the direction we expect, namely higher $\Omega_c h^2$, higher $A_s$, lower $n_s$, and lower $H_0$~\cite{Planck2018Parameters}. 

This extra smoothing results in a PTE of $\ExtraSmoothApteparamshift$ ($\ExtraSmoothAsnrparamshift$) comparing the shift in best-fit parameters between lensed and delensed spectra over 20 free parameters. We see that this deviation is coming from the $\Lambda $CDM block of the parameter-shift covariance matrix, with a PTE of $\ExtraSmoothApteparamshiftLCDMBlock$ ($\ExtraSmoothAsnrparamshiftLCDMBlock$ for 6 dof) marginalizing over the 14 foreground and calibration parameters. In contrast, the PTE marginalizing over the 6 $\Lambda $CDM parameters is $\ExtraSmoothApteparamshiftFGBlock$ ($\ExtraSmoothAsnrparamshiftFGBlock$ for 14 dof).  If we increase the extra smoothing to $30\%$, i.e.~make $A_L=1.3$, then the parameter-shift test deviates at $\ExtraSmoothBsnrparamshift$ with a PTE of $\ExtraSmoothBpteparamshift$ (20 dof).  Marginalizing over the foreground plus calibration factors results in a PTE of $\ExtraSmoothBpteparamshiftLCDMBlock$ ($\ExtraSmoothBsnrparamshiftLCDMBlock$ for 6 dof).  In contrast, the PTE marginalizing over the $\Lambda $CDM parameters is $\ExtraSmoothBpteparamshiftFGBlock$ ($\ExtraSmoothBsnrparamshiftFGBlock$ for 14 dof).  This demonstrates the ability of this parameter-shift statistic not only to detect a deviation from the $\Lambda $CDM plus foreground model, but also to identify whether the deviation is due to the foreground model or the $\Lambda $CDM cosmological model. This ability arises essentially because foregrounds do not mimic acoustic peak smearing, as is well known. Repeating the above on 10 different simulation realizations with $A_L=1.2$ and $A_L=1.3$ we find, marginalizing over foregrounds, a mean PTE of $\MeanExtraSmoothApteparamshiftLCDMBlock$ and $\MeanExtraSmoothBpteparamshiftLCDMBlock$ respectively, corresponding to $\MeanExtraSmoothAsnrparamshiftLCDMBlock$ and $\MeanExtraSmoothBsnrparamshiftLCDMBlock$ deviations (for 6 dof) respectively (see Figure~\ref{fig:AL} in Appendix~\ref{sec:appendixParamShiftCov}).

To explain more intuitively why this parameter-shift statistic provides a novel lensing consistency test, let us consider the case where delensing removes $30\%$ of the lensing signal in the CMB power spectrum (as in this work). Let us also assume that there is extra lensing-like peak smoothing in the CMB power spectrum.
Let the true amount of lensing be $X$, and let us measure $X+E$, where $E$ is the extra smoothing in the CMB spectrum.  By internally reconstructing the lensing signal from CMB measurements, we can directly measure the true lensing signal $X$. When we delens, we remove $0.3~X$; however, we will measure $Y = 0.7~X + E$ (the spurious lensing-like signal $E$ will not be affected by the delensing procedure). Thus the  expected ratio of lensing in the delensed and lensed power spectra should be $0.7$, but instead will be $(0.7~X + E)/(X+E)$.  (Note that the more one can delens, the larger will be this discrepancy between expected and measured ratios.)  When we fit the delensed spectra, we will then obtain a set of inferred $\Lambda$CDM parameters that will differ from the set obtained from the lensed spectra.  

This parameter-shift statistic is a more sensitive probe of inconsistent lensing in the CMB power spectrum than fitting to an $A_L$ parameter as is done in A20.{\color{blue}\footnote{Note A20 uses the full ACT DR4 data set which has more than twice the data presented here.}}  This is because this parameter-shift test directly probes the consistency between lensing-induced peak smoothing and the reconstructed lensing map, and does not suffer from sample variance. Thus, going forward with future CMB datasets, the parameter-shift statistic we introduce here can provide a novel way to search for physics in the early Universe that differs from $\Lambda$CDM.

\subsection{Parameter Results from ACT Data}
\label{sec:results}
We unblind the data following the procedure outlined in the steps below. 
\begin{enumerate}
  \item Delensing pipeline and likelihood code are verified as described in Section~\ref{sec:pipe} and Appendix~\ref{sec:appendixParamShiftCov}.
  \item A set of null tests is performed on the data. For details of these null tests we refer to~C20.
  \item Without unblinding the cosmological and foreground parameter results, we calculate the reduced $\chi^2$ and the PTEs of the best-fit point in the CosmoMC chains for both lensed and delensed spectra.  CosmoMC returns the $\chi^2$ of this best-fit point, and we assume $490-22=468$ degrees of freedom (since we have 10 spectra, 49 bins per spectra, and 22 free parameters).{\color{blue}\footnote{Note that after unblinding we updated the $TT~\ell_{\rm{min}}$ from 475 to 575 and removed the $A^{TE}_{sync}$ and $A^{EE}_{sync}$ foreground parameters, as discussed later in this section.  This reduces the number of $TT$ spectral bins to 47, and changes the total degrees of freedom to $484-20=464$.}} We check whether the PTE values are within a reasonable range ($0.05 \leq PTE \leq 0.95$).
  \item We unblind the marginalized mean parameter results for the lensed and delensed data spectra obtained from the CosmoMC chains.
  \item We generate theory curves derived from the marginalized mean parameters for the lensed and delensed spectra. Given these theory curves, we calculate the delensing signal-to-noise ratio, the delensing efficiency, and the PTEs of the delensed spectra with respect to the delensed theory curves.
  \item Then we run the $action=2$ statistic described in Section~\ref{sec:param-shift-covmat} to find the best-fit values for the lensed and delensed spectra.  
  \item We then difference these best-fit values obtained for the lensed and delensed spectra, and calculate the resulting PTE, using the parameter-shift covariance matrix, to determine whether it is within a reasonable range.
  
\end{enumerate}

\noindent Following the unblinding protocols enumerated above, we unblind the data and obtain the cosmological parameter results presented in Appendix~\ref{sec:appendixOrigResults}.\\

After unblinding the data, we make the following changes based on additional information, as is also done in~C20 and~A20. 
All of the sections in this paper that are presented above incorporate these changes.  We present the initial unblinded parameters in Figure~\ref{fig:compare_data} of Appendix~\ref{sec:appendixOrigResults}. 
\begin{enumerate}
\item We find after unblinding the data that the Galactic synchrotron parameters, $A^{TE}_{sync}$ and $A^{EE}_{sync}$ are consistent with zero. 
In addition, there is no evidence for Galactic synchrotron emission in D56 as presented in~C20.  Thus we remove these two foreground parameters from our likelihood and fix their values to zero as is also done in~C20. This gives us 14 foreground parameters as opposed to our original 16. 
\item We limit the range of the Poisson amplitude of polarized radio sources in the $EE$ spectrum, $A^{EE}_{PS}$, to be positive after initially finding a result consistent with zero.
\item We update the priors on the Galactic dust levels based on analysis presented in~C20 cross correlating 353 GHz {\it{Planck}} data with 150 GHz ACT DR4 deep patch data.  Since the D56 sky region used in this work carries most of the DR4 deep patch weight, we use this information to add priors on the Galactic dust parameters, $A^{TT}_{dust,d}$, $A^{TE}_{dust,d}$, and $A^{EE}_{dust,d}$, shown in Table~\ref{tab:priors}. 
\item We add a prior on the clustered part of the cosmic infrared background (CIB) of $4.9 \pm 0.9$ based on previous ACT data that included a 220 GHz channel~\cite{Dunkley2013}.  
\item We shift the prior on the Poisson part of the CIB to be $3.1 \pm 0.4$ instead of $2.9 \pm 0.4$, and shorten the prior range of the CIB spectral index, $\beta_c$, from [0, 8] to [0, 5] to be consistent with~C20.
\item We correct a bug in the calibration factors for the data that mainly affected the 98 GHz channel; this correction shifted the 98 GHz spectra up by $3\%$. 
\item We increase the $\ell_{\rm{min}}$ for the $TT$ spectra to be 575 instead of our original 475, based on information gained by cross correlating with {\it{Planck}} data and possible systematics at lower $\ell$ in our $TT$ data, as presented in~C20.
\item We correct for residual temperature to polarization leakage and polarized beam buddies described in~A20 following the procedure in~C20. We find that these corrections result in changes of less than $0.6\%$ to the cosmological parameter results presented in this work.  
\end{enumerate}

\begin{figure*}[t]
  \centering
  \includegraphics[width=1.0\textwidth]{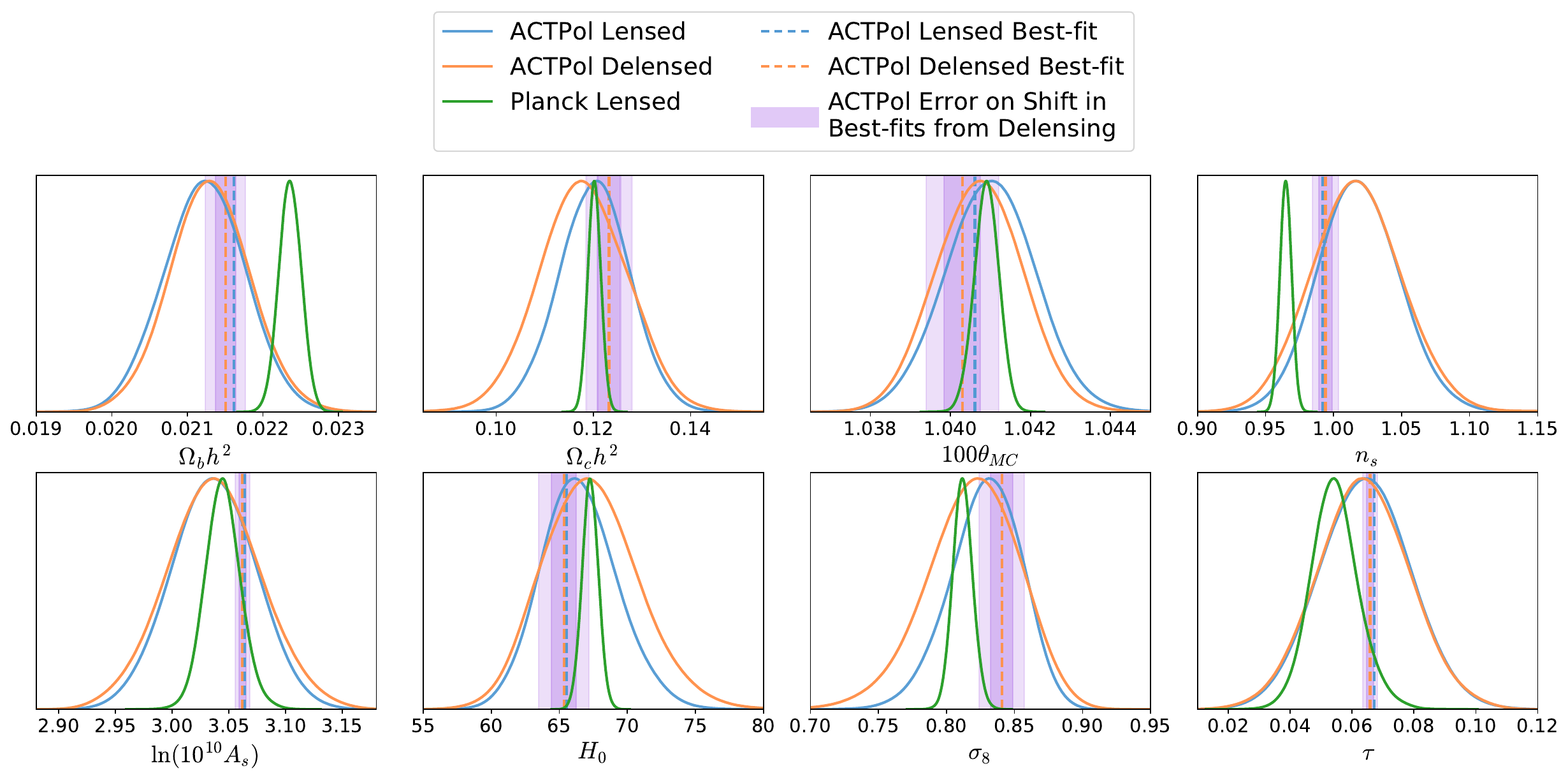}
  \caption{Shown are the marginalized mean $\Lambda$CDM parameter constraints from ACT D56 lensed and delensed data power spectra (blue and orange curves). We also show the marginalized mean parameters from {\it{Planck}} lensed power spectra (green curves)~\cite{Planck2018Parameters}. As discussed in C20, $\Omega_b h^2$ and $n_s$ for ACT versus {\it{Planck}} become more consistent with the inclusion of external data to constrain the amplitude of the first peak of the $TT$ spectrum. The dashed blue and orange vertical lines indicate the best-fit parameters from ACT lensed and delensed spectra. (Note for $\Omega_c h^2$ and $\sigma_8$ the dashed blue line is under the orange line.) These marginalized mean and best-fit parameters are given in Table~\ref{tab:params}. The purple solid bands (centered on the best-fit parameters from the delensed spectra) indicate the $1\sigma$ and $2\sigma$ error on the shift in best-fit parameters between ACT lensed and delensed spectra obtained from the diagonal elements of the parameter-shift covariance matrix described in Section~\ref{sec:param-shift-covmat}.  We see that the allowed uncertainty on a shift in best-fit parameters is much smaller than the error on each parameter individually due to sample-variance cancellation. A separation in best-fit values wider than the purple bands would indicate a deviation from the  $\Lambda$CDM model or an unknown systematic effect.}
  \label{fig:params}
\end{figure*}

\begin{table*}[t]
\begin{center}

\begin{tabular}{c @{\hskip 0.02\textwidth} c @{\hskip 0.02\textwidth} c  @{\hskip 0.025\textwidth} c @{\hskip 0.02\textwidth} c @{\hskip 0.025\textwidth} c @{\hskip 0.025\textwidth} c}

\hline
\hline

& \multicolumn{2}{c}{ACT-D56 Lensed} & \multicolumn{2}{c}{ACT-D56 Delensed} &  \multicolumn{1}{c}{\hspace{-2em} {ACT Full DR4 Lensed}}  & \multicolumn{1}{c}{{\it{Planck}} Lensed} \\ 
Parameter & Marginalized Mean & Best Fit & Marginalized Mean & Best Fit & Marginalized Mean & Marginalized Mean\\

\hline 

$\Omega_b h^2$ \dotfill \rule{0pt}{2.9ex} & 0.02136 $\pm$ 0.00055  & 0.02161  & 0.02130 $\pm$ 0.00053 & 0.02150 & 0.02154  $\pm$ 0.00030  & 0.02236 $\pm$ 0.00015 \\ 
$\Omega_c h^2$ \dotfill                   & 0.1199 $\pm$ 0.0072    & 0.1232 & 0.1179  $\pm$ 0.0091  &  0.1232   & 0.1177  $\pm$ 0.0038  & 0.1202 $\pm$ 0.0014 \\
$100 \theta_{MC}$ \dotfill                & 1.0410 $\pm$ 0.0011    & 1.0406  & 1.0407  $\pm$ 0.0011  & 1.0403   & 1.04225 $\pm$ 0.00072  & 1.04090 $\pm$ 0.00031 \\
$\tau$ \dotfill                           & 0.064 $\pm$ 0.015      &  0.067 & 0.064   $\pm$ 0.015   & 0.066     & 0.065 $\pm$ 0.014  & 0.0544$^{+0.0070}_{-0.0081}$ \\
$n_s$ \dotfill                            & 1.013 $\pm$ 0.029      &  0.992 & 1.017   $\pm$ 0.033   &  0.994    & 1.008 $\pm$ 0.016 & 0.9649 $\pm$ 0.0044  \\
$\ln(10^{10} A_s)$ \dotfill               & 3.040 $\pm$ 0.036      &  3.064 & 3.037   $\pm$ 0.042   &  3.062    & 3.050 $\pm$ 0.030  & 3.045 $\pm$ 0.016 \\
\hline
$\sigma_8$ \dotfill                       & 0.827 $\pm$ 0.026      &  0.841 & 0.819   $\pm$ 0.034     & 0.841   & 0.824 $\pm$ 0.016  & 0.8120 $\pm$ 0.0073 \\
$H_0$ \dotfill \rule{0pt}{2.9ex}          & 66.7 $\pm$ 2.8         &  65.5 & 67.3   $\pm$ 3.5    &  65.3    & 67.9  $\pm$ 1.5  & 67.27 $\pm$ 0.60 \\
\hline

\end{tabular}
\caption{Shown are the marginalized mean and best-fit $\Lambda$CDM parameter values from ACT lensed and delensed power spectra presented in this work. The first six rows contain the cosmological parameters that are sampled during the cosmological analysis, and the last two rows contain the derived parameters $\sigma_8$ and $H_0$. Included in the last two columns for comparison are the marginalized mean parameter values from the ACT Full DR4 given in~A20 and the {\it{Planck}} {\it{TT,TE,EE+lowE}}~\cite{Planck2018Parameters} lensed spectra.  As described in C20, combining ACT DR4 with {\it{WMAP}} reduces the tension of $\Omega_b h^2$ and $n_s$ compared to {\it{Planck}}.}
\label{tab:params}
\end{center}
\end{table*}

\begin{figure*}[t]
  \centering
  \includegraphics[width=1.0\textwidth]{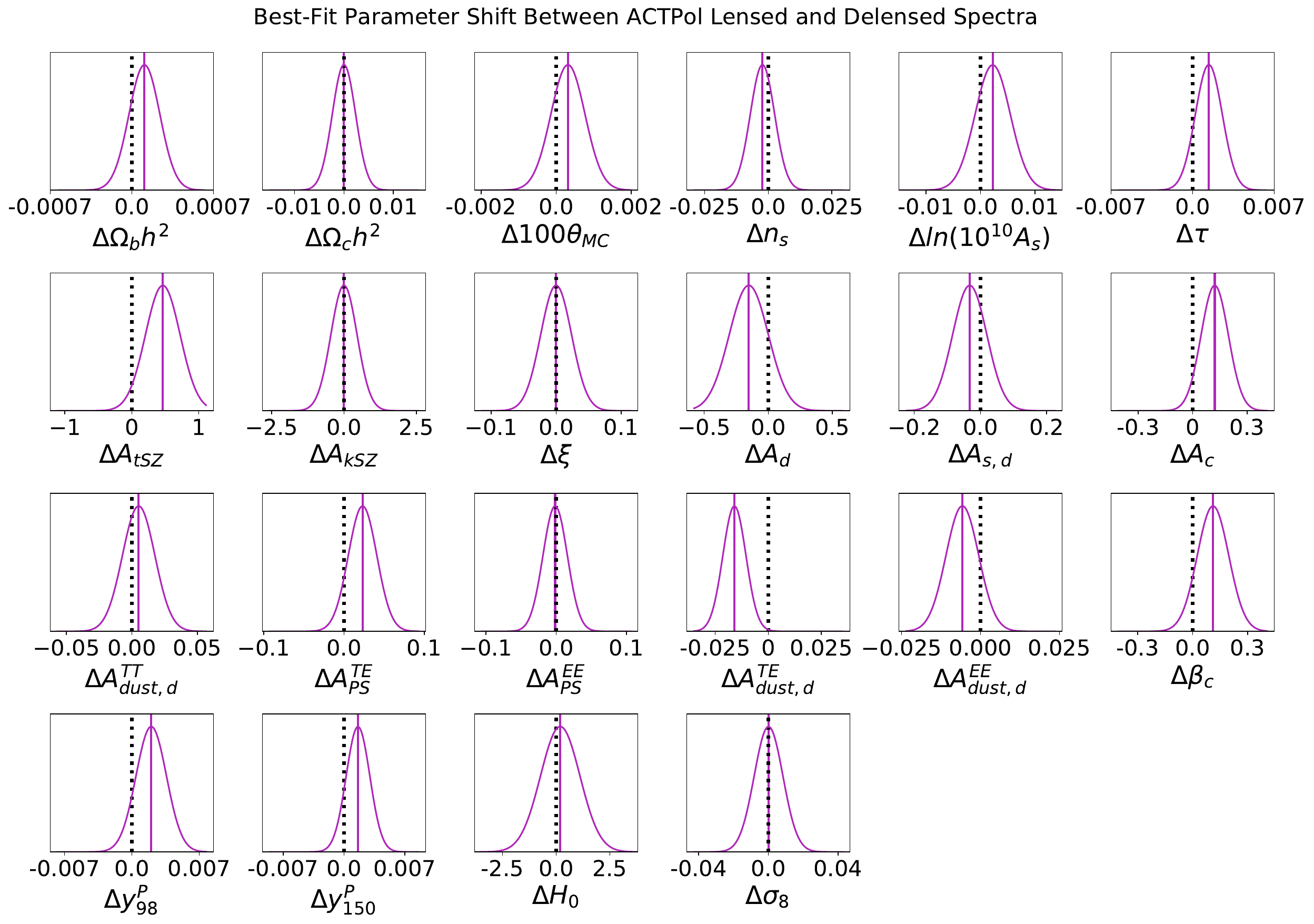}
  \caption{The solid purple vertical lines show the best-fit parameters from ACT lensed spectra minus the best-fit parameters from ACT delensed spectra (i.e.~the shift in best-fit parameters). The marginalized uncertainty on the shift in each parameter is obtained from the diagonal elements of the parameter-shift covariance matrix described in Section~\ref{sec:param-shift-covmat}; this uncertainty is indicated by the solid purple curves, which are a zoom-in of the purple bands shown in Figure~\ref{fig:params}.  The dotted black lines indicate zero shift in a parameter. We also add the shift in the derived parameters, $H_0$ and $\sigma_8$, for completeness.  Using the full 20x20 parameter-shift covariance matrix, we find a shift in parameters that deviates from the expectation of our $\Lambda $CDM plus foreground model at $\snrparamshift$.  In this figure, we can see that some of the largest shifts are from polarized Galactic foregrounds, $\Delta A^{TE}_{dust,d}$ and $\Delta A^{EE}_{dust,d}$, which are in the data but whose scatter is not modeled in our simulations (see Section~\ref{sec:results} for more detail).   Marginalizing over foregrounds, we find that the shift in $\Lambda$CDM parameters agrees with expectations within $\snrparamshiftLCDMBlock$.}
  \label{fig:paramShift}
\end{figure*}

Table~\ref{tab:params} and Figure~\ref{fig:params} give the marginalized mean $\Lambda$CDM cosmological parameter constraints from ACT lensed and delensed spectra (blue and orange solid curves in the figure).  They also give the best-fit parameters for ACT lensed and delensed spectra, obtained from averaging twenty $action=2$ runs (blue and orange dashed vertical lines in the figure). Figure~\ref{fig:params} also shows via purple solid bands the 1 and 2$\sigma$ error on the shift in best-fit parameters between ACT lensed and delensed spectra; these errors we obtain from the diagonal elements of the parameter-shift covariance matrix described in Section~\ref{sec:param-shift-covmat}.  As is shown, the allowed uncertainty on a shift in best-fit parameters is much smaller than the error on each parameter individually due to sample-variance cancellation. We also include in this figure the marginalized mean parameters from {\it{Planck}} lensed spectra calculated over the full sky (green solid curves)~\cite{Planck2018Parameters}.  We obtain these marginalized mean parameters from the available chains in the {\it{Planck}} legacy archive.  We also confirm these parameters by  using CosmoMC and the latest 2018 \textit{Planck} data and likelihood~\cite{Planck2018likelihood, Planck2018Parameters}; we run eight CosmoMC chains using the \textit{Planck} 2018 baseline \textit{TT,TE,EE+lowE} data, and the default CosmoMC flat priors on the six varied cosmological parameters.  

We find that the marginalized mean parameters from both ACT lensed and delensed spectra are consistent with each other and with the marginalized mean parameters obtained from {\it{Planck}}.  For the delensed spectra, the model from the marginalized mean parameters yields a reduced $\chi^2$ of $\redChiALL$ (PTE= $\pteALL$) as shown in Table~\ref{tab:data}. The derived parameters $H_0$ and $\sigma_8$ obtained from delensed~$TT, TE,$ and $EE$ spectra are: 
\begin{align}
H_0&=\hubbleConstraint &{\rm{ACT~D56}}\\
\sigma_8&= \sigmaEightConstraint &{\rm{ACT~D56}}
\end{align}
From Table~\ref{tab:params} we see that the error bars are larger on some of the parameters obtained from the delensed spectra compared to the lensed spectra.  In order to obtain tighter parameter constraints from delensing CMB power spectra one needs to combine that information with additional information from the CMB lensing power spectrum~\cite{Green2017}.  This is because while delensing sharpens the acoustic peaks and improves cosmological parameters that affect the acoustic structure, delensing also removes lensing information in the CMB power spectrum; if the lensing power spectrum were used together with the delensed CMB power spectra, which is beyond the scope of this paper, this lost lensing information would be added back~\cite{Green2017}.  

Figure~\ref{fig:params} also shows that the shifts in the best-fit $\Lambda$CDM parameters (dashed vertical lines) are within the expected scatter (shown by the vertical bands).  To quantify this, we marginalize over foreground and calibration parameters (i.e.,~we compute the $\chi^2$ of  Eq.~\ref{eq:chi2paramshift} using only the 6x6 block of $\Lambda $CDM parameters, corresponding to the top-left corner of Figure~\ref{fig:full-param-shift-covmat}); doing this we obtain a PTE of~$\pteparamshiftLCDMBlock$, corresponding to only a~$\snrparamshiftLCDMBlock$ deviation from expectation (6 dof).

Figure~\ref{fig:paramShift} is a zoom-in on the difference between best-fit parameters derived from ACT lensed and delensed spectra, now including the foreground and calibration parameters.  Here a dotted vertical line indicates zero shift between best-fit lensed and delensed parameters. The shift in best-fit parameters that we measure is given by the solid purple line, and the allowed marginalized uncertainty based on the diagonals of the parameter-shift covariance matrix is given by the solid purple curves. 

In many CMB analyses one can separate $\Lambda$CDM parameters from foreground and calibration parameters, and the same is true for the parameter-shift statistic.  While the parameter-shift statistic does not inform the mean values or error bars of parameters, for example, as determined in C20 and A20, it can inform whether the correlated shift in parameters matches expectations. 
We find a PTE of~$\pteparamshift$ using the measured parameter shifts shown in Figure~\ref{fig:paramShift} and the full 20x20 parameter-shift covariance matrix (depicted in Figure~\ref{fig:full-param-shift-covmat}). This PTE corresponds to a~$\snrparamshift$ deviation from the expectation of our $\Lambda $CDM plus foreground model, given 20 degrees of freedom.  When we marginalize over the $\Lambda $CDM parameters (i.e.~use the bottom right 14x14 foreground and calibration block of the covariance matrix), we obtain a PTE of~\pteparamshiftFGBlock, corresponding to a~$\snrparamshiftFGBlock$ deviation from expectations (14 dof). This indicates that the deviation we see is from the foreground plus calibration model, and not from the $\Lambda$CDM parameters.  Note that this deviation from expectation would not have impacted the C20 and A20 CMB power spectra analyses; it is a feature that is uniquely relevant here.

When we break down the foreground plus calibration sector into smaller pieces and marginalize over the other parameters, we find that the largest deviations from expectation are coming from the modelling of the extragalactic temperature foregrounds ($\snrparamshiftExtraGalTemp$ deviation for 7 dof), and the polarized Galactic foregrounds ($\snrparamshiftGalPol$ deviation for 2 dof), with the combination of these yielding a~$\snrparamshiftGalPolExtraGalTemp$ deviation (9 dof) from our model expectation.  All of the other foreground plus calibration factors combined yield an acceptable PTE ($\snrparamshiftFGRest$ for 5 dof).  

The deviation due to the polarized Galactic foregrounds ($\snrparamshiftGalPol$) is likely because we do not model polarized Galactic foregrounds in our simulations described in Section~\ref{sec:sims}.  Thus our parameter-shift covariance matrix is not capturing the scatter from polarized Galactic dust nor the correlations between $A^{\rm{TE}}_{\rm{dust,d}}$ and $A^{\rm{EE}}_{\rm{dust,d}}$, as seen in Figure~\ref{fig:full-param-shift-covmat}.

When we break down the extragalactic temperature foregrounds (that had a deviation of $\snrparamshiftExtraGalTemp$) into smaller subsets, we find that the largest deviation from expectation is coming from the amplitude of the kinetic Sunyaev-Zel'dovich (kSZ) effect ($A_{\rm{kSZ}}$) and the cross-correlation between the thermal Sunyaev-Zel'dovich effect and the cosmic infrared background ($\xi$); these two parameters alone give a PTE of $\pteparamshiftKSZXI$ (corresponding to $\snrparamshiftKSZXI$ for 2 dof).  Marginalizing over these two parameters, the other extragalactic temperature foregrounds yield an acceptable PTE of $\pteparamshiftExtraGalTempRest$~($\snrparamshiftExtraGalTempRest$ for 5 dof).  We find that the best-fit values of these foreground parameters have a preference for $A_{\rm{kSZ}}=0$ and $\xi=0.2$, both of which are at the boundaries of their prior ranges (see Table~\ref{tab:priors}).  In Figure~\ref{fig:paramShift}, we see zero shift in these parameters between lensed and delensed data spectra, whereas we do see shifts in the other extragalactic temperature foreground parameters.  This is not expected given our parameter-shift covariance matrix, shown in Figure~\ref{fig:full-param-shift-covmat}, which shows significant correlations between all the extragalactic temperature foreground parameters.  This suggests that more refinement of the modelling of $A_{\rm{kSZ}}$ and $\xi$ may be warranted to better match the data.  

We also further check that there is no correlation between the lensed minus delensed CMB map, shown in Figure~\ref{fig:LensedMinusDelensed}, and Galactic dust in temperature, in case, for example, there is residual Galactic dust in our lensing potential map.  We cross correlate the \textit{Planck} 545 GHz Galactic dust temperature map obtained from~\cite{Planck2015Component} with the lensed minus delensed map from the data and from 512 simulations.  We use the cross correlation with the simulations to obtain the covariance.  We find a cross correlation consistent with zero, with PTEs  of $\ptedustcrossPlanckHunderFifty$ and $\ptedustcrossPlanckNinty$ for the 150 and 98 GHz cross correlations respectively.

\section{Discussion}
\label{sec:discussion}

We have presented for the first time cosmological parameters obtained from delensed CMB power spectra.  
The combination of 150 and 98~GHz $TT, EE,$ and $TE$ delensed spectra from ACT data covering 482 square degrees of sky are well fit by a standard $\Lambda $CDM model. We find marginalized mean parameters from lensed and delensed spectra that are consistent with each other and with the latest {\it{Planck}} 2018 results~\cite{Planck2018Parameters}.  

We have also presented a new tool -- the shift in parameters between lensed and delensed spectra -- that can allow us to explore the match between data and model in a different way than standard techniques to date.  Marginalizing over foreground and calibration parameters, we find that the shift in $\Lambda$CDM parameters is consistent with zero within $\snrparamshiftLCDMBlock$.  This implies that the lensing in the CMB power spectrum is consistent with expectations. To put this result in context, as discussed in Section~\ref{sec:newPhysics} and shown in Figure~\ref{fig:AL}, if the effective peak-broadening parameter $A_L$ had an actual value of 1.2, our shift statistic with the current ACT data would detect the difference from the $\Lambda$CDM value of $A_L=1$ at a statistical significance of 2$\sigma$.  Our result is also consistent with the $A_L$ constraint from~A20 using the full ACT DR4 data set.

Upcoming microwave background temperature and polarization maps with higher sensitivity from the Advanced ACT Polarimeter, as well as from future experiments such as the Simons Observatory~\cite{SO2019}, CMB-S4~\cite{CMB-S4-ScienceBook}, and CMB-HD~\cite{CMB-HD-WhitePaper, CMB-HD-RFI}, hold the possibility of delensing with significantly improved efficiency. The powerful consistency tests that result will provide a novel diagnostic of possible systematic errors in these challenging systematics-dominated experiments. Ultimately, internal consistency tests of cosmological models from microwave background data alone offer the enticing possibility of revealing any departures from the standard cosmological model which our universe may hold.

\begin{figure*}[t]
  \centering
  \includegraphics[width=0.9\textwidth]{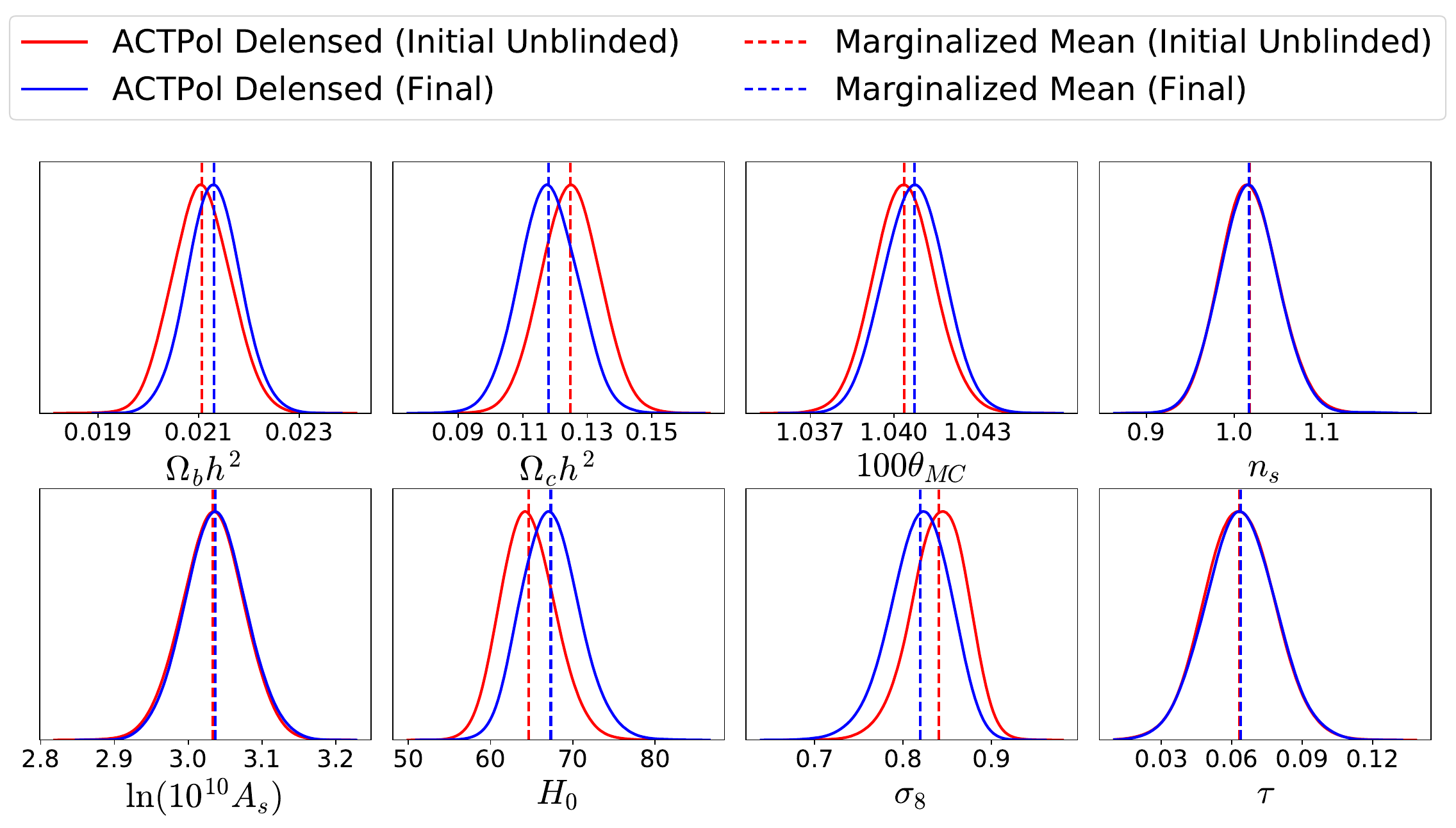}
  \caption{Comparison of the cosmological parameters from delensed spectra obtained just after unblinding and the final cosmological parameters presented in this work.  See Section~\ref{sec:results} for details about the changes made post unblinding.}
  \label{fig:compare_data}
\end{figure*}

\begin{acknowledgments}
The authors would like to thank Joel Meyers for discussion regarding modifications to the theory code in Ref~\cite{Green2017} that we performed.  DH, NS, and AM acknowledge support from NSF grant numbers AST-1513618 and AST-1907657.  EC acknowledges support from the STFC Ernest Rutherford Fellowship ST/M004856/2 and STFC Consolidated Grant ST/S00033X/1, and from the Horizon 2020 ERC Starting Grant (Grant agreement No 849169).  MSM acknowledges support from NSF grant AST-1814971.  KM and MH acknowledges support from the National Research Foundation of South Africa.  This work was supported by the U.S.~National Science Foundation through awards AST-0408698, AST-0965625, and AST-1440226 for the ACT project, as well as awards PHY-0855887 and PHY-1214379. Funding was also provided by Princeton University, the University of Pennsylvania, and a Canada Foundation for Innovation (CFI) award to UBC. ACT operates in the Parque Astronómico Atacama in northern Chile under the auspices of the Comisión Nacional de Investigación Científica y Tecnológica de Chile (CONICYT). Computations were performed on the GPC supercomputer at the SciNet HPC Consortium. SciNet is funded by the CFI under the auspices of Compute Canada, the Government of Ontario, the Ontario Research Fund - Research Excellence; and the University of Toronto. This research also used resources of the National Energy Research Scientific Computing Center, which is supported by the Office of Science of the U.S. Department of Energy under Contract No.~DE-AC02-05CH11231.  In addition, computations were performed on the supercomputers provided by the Center for Computational Astrophysics at the Flatiron Institute. The Flatiron Institute is supported by the Simons Foundation.  The development of multichroic detectors and lenses was supported by NASA grants NNX13AE56G and NNX14AB58G. We thank our many colleagues from ABS, ALMA, APEX, and Polarbear who have helped us at critical junctures. Colleagues at AstroNorte and RadioSky provide logistical support and keep operations in Chile running smoothly. We also thank the Mishrahi Fund and the Wilkinson Fund for their generous support of the project.
\end{acknowledgments}

\appendix

\section{Initial Unblinded Results}
\label{sec:appendixOrigResults}

As discussed in Section~\ref{sec:results}, we initially performed the analysis with the parameter priors listed in Table~\ref{tab:original_priors}.  We also originally used $\ell_{\rm{min}}=475$, as opposed to $\ell_{\rm{min}}=575$, for the $TT$ power spectra. When we unblinded the data we obtained from delensed~$TT, TE,$ and $EE$ spectra: 
\begin{align}
H_0&=\hubbleConstraintOrig  &{\rm{ACT~D56}}\\
\sigma_8&= \sigmaEightConstraintOrig  &{\rm{ACT~D56}}
\end{align}
As shown in Figure~\ref{fig:compare_data}, there is small change in the cosmological parameter results after making the updates to the likelihood and analysis discussed in Section~\ref{sec:results}. We also initially obtained a PTE of $\pteparamshiftOrig$ using the measured parameter shifts between lensed and delensed spectra, using the original 22x22 parameter-shift covariance matrix.  This PTE corresponds to a $\snrparamshiftOrig$ deviation from the expectation of our $\Lambda $CDM plus foreground model, assuming 22 degrees of freedom.  The final PTE for the parameter shifts we obtain is $\pteparamshift$ which is a $\snrparamshift$ deviation from the expectation and similar to the initial unblinded value.

\begin{table}[h]
\begin{tabularx}{0.45\textwidth}{|cc|cc|}
\cline{1-4}
~~Parameter    ~~       & Prior    ~~~~     & Parameter           & Prior          \\
\cline{1-4}
~~$A_{s,d}$  ~~   & $2.9 \pm 0.4$ ~~~~ & $A_c$ & [0, 15] ~~~\\
~~$A^{TE}_{PS}$  ~~   & [-1, 1]  ~~~~   &  $\beta_c$           & [0, 8]  \\ 
~~$A^{EE}_{PS}$ ~~  &  [-2, 2]   ~~~~  & $A_d$     & [0, 11] \\
~~$A^{TT}_{dust,d}$  ~~   & [-20, 20]  ~~~~ &~~~  ~~$A_{tSZ}, A_{kSZ}$ & [0, 10] \\
~~$A^{TE}_{dust,d}$  ~~   & [-3, 3] ~~~~ &  $\xi$   & [0, 0.2] \\
~~$A^{EE}_{dust,d}$  ~~   &  [-2, 2] ~~~~ & $y^{P}_{98},y^{P}_{150}$ & [0.9, 1.1] ~~\\ 
\rule[-1.2ex]{0pt}{0pt} ~~ $A^{TE}_{synch}$, $A^{EE}_{synch}$  ~~   &  [-2, 2] ~~~~ & &  \\ 
\cline{1-4}
\end{tabularx}
\caption{Shown are the initial prior ranges for the 16 foreground parameters used in the parameter analysis before unblinding.}
\label{tab:original_priors}
\end{table}

\section{Parameter-Shift Covariance Matrix}
\label{sec:appendixParamShiftCov}

In this appendix, we discuss the properties of the parameter-shift covariance matrix, discussed in Section~\ref{sec:param-shift-covmat}, in more detail. \\

\begin{figure}[t]
  \centering
  \includegraphics[width=0.49\textwidth]{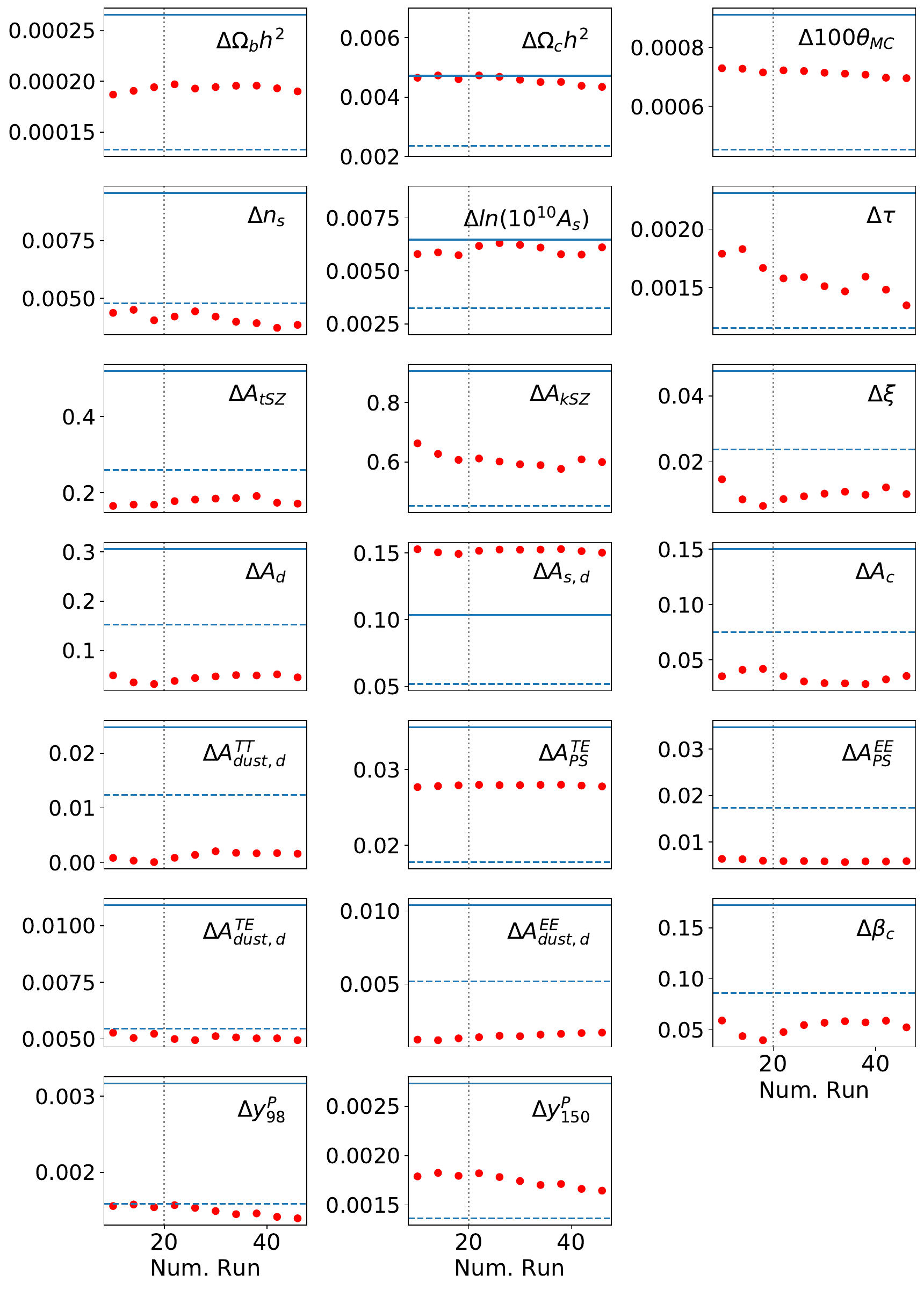}
  \caption{Shown are the running averages of best-fit parameter-shift values as a function of $N$ $action=2$ CosmoMC runs, derived from a single simulated lensed/delensed spectra pair (red dots).  The dashed and solid blue curves show for reference the $1\sigma$ and $2\sigma$ diagonal errors of the parameter-shift covariance matrix shown in Figure~\ref{fig:full-param-shift-covmat}.  We find the running averages become stable after roughly ten $action=2$ runs. For this analysis, we decide to average over twenty $action=2$ runs (dotted vertical lines) to make sure we have converged shifts.
  }
  \label{fig:action=2Convergence}
\end{figure}

\begin{figure}[t]
  \centering
  \includegraphics[width=0.49\textwidth]{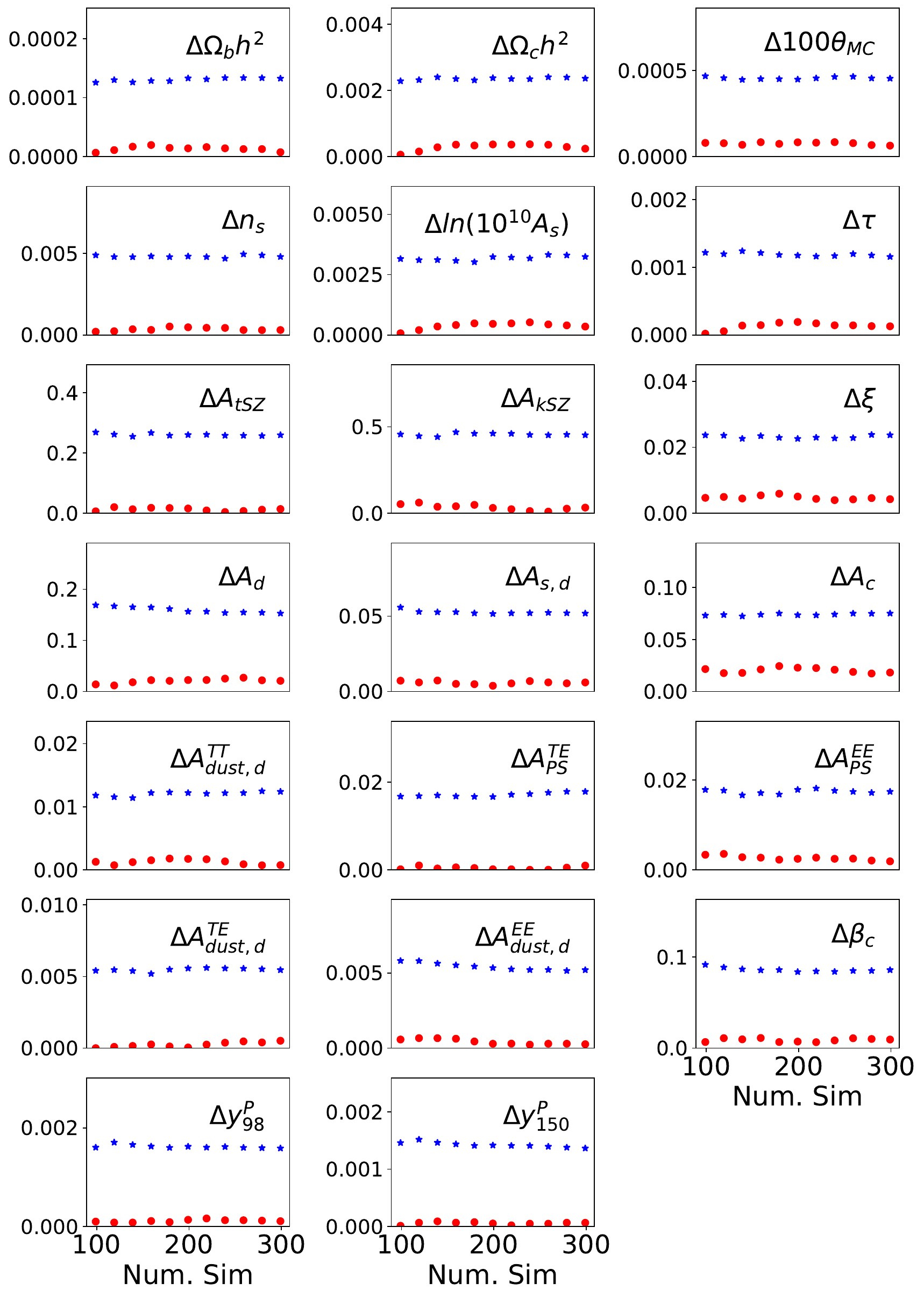}
  \caption{We show the convergence of the parameter-shift covariance matrix diagonal elements as a function of $N$ simulations.  The absolute value of the mean parameter shift converges to zero (red dots), and the scatter is well converged after 300 simulations (blue stars).}
  \label{fig:ParamShiftConvergence}
\end{figure}

\begin{figure*}[!htbp]
 \centering
  \includegraphics[width=1\textwidth]{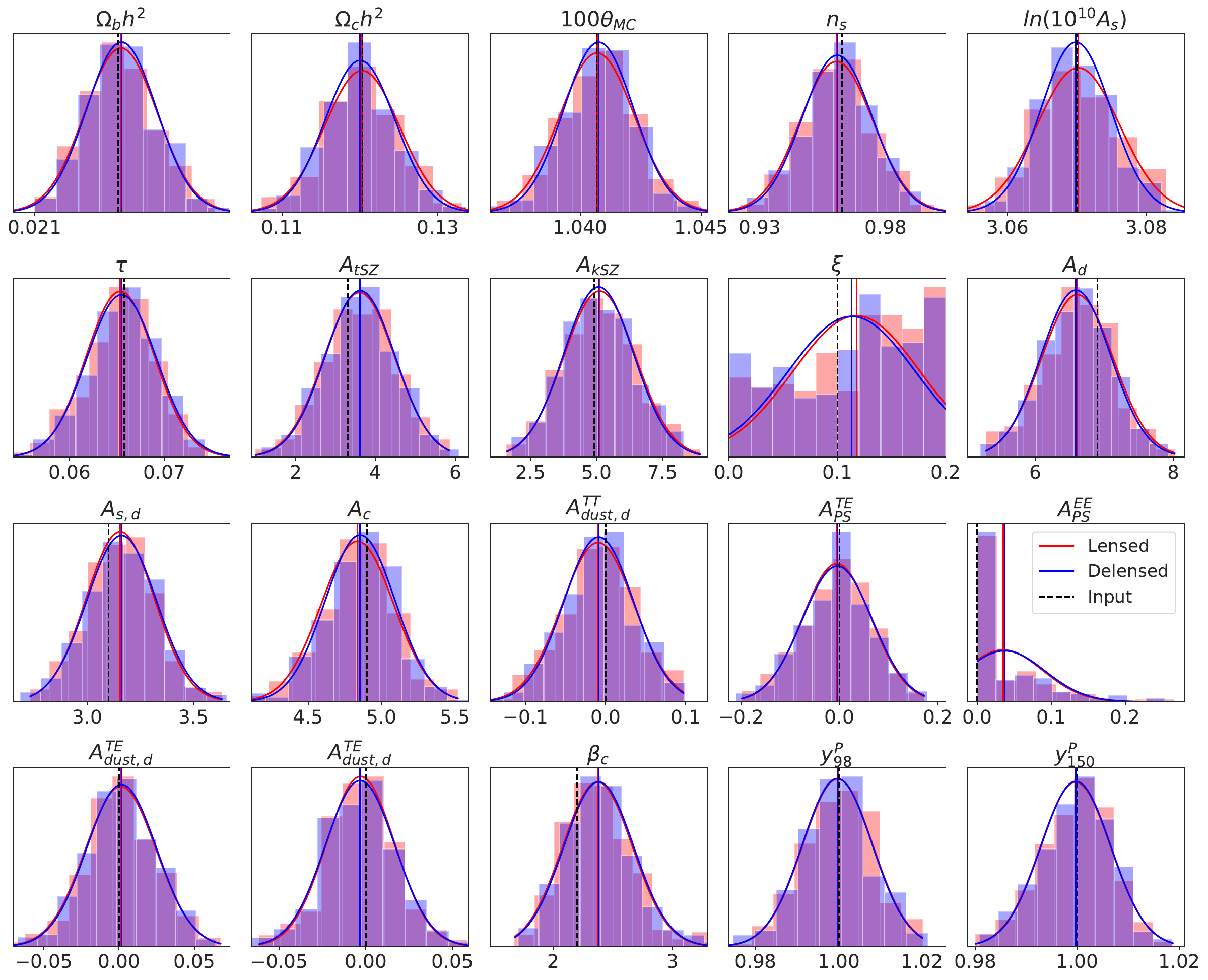}
  \caption{Shown is the agreement between best-fit parameters obtained from lensed (red histograms) and delensed (blue histograms) spectra from 300 simulations in simulation $set_3$, which is discussed in Section~\ref{sec:sims}. The red and blue curves are Gaussian distributions centered on the means of the best-fit parameters obtained from the lensed and delensed spectra, respectively, which are indicated by the red and blue vertical lines. The widths of these distributions are determined by the scatter in the best-fit parameter values. The dashed black vertical lines show the input parameters of the simulations, indicating no significant bias in the best-fit parameters}.
  \label{fig:ParamShiftBias1}
\end{figure*}

\begin{figure*}[!htbp]
 \centering
  \includegraphics[width=1\textwidth]{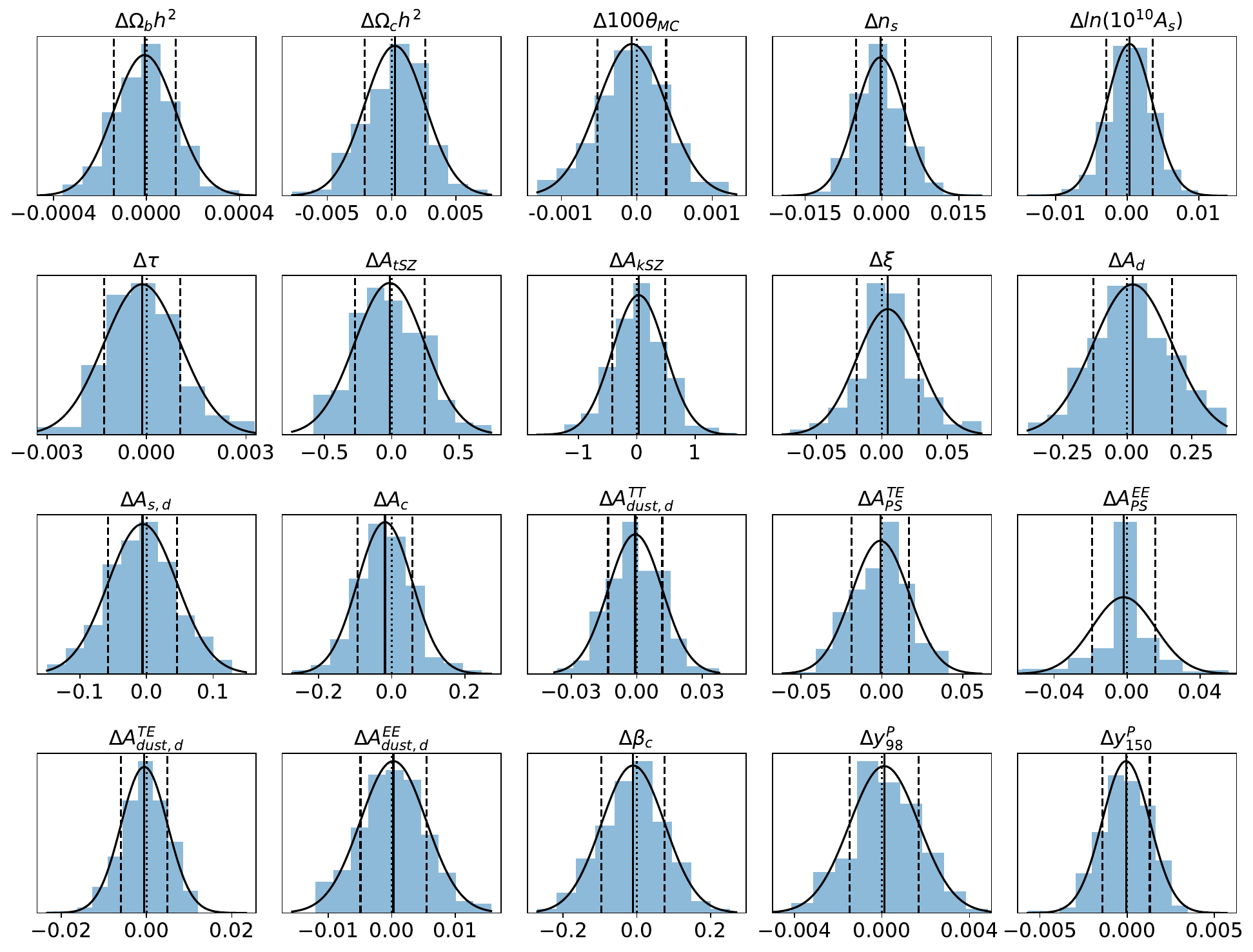}
  \caption{Shown is the distribution of parameter shifts (blue histograms) derived from lensed versus delensed spectra for the 300 simulations shown in Figure~\ref{fig:ParamShiftBias1} and described in Section~\ref{sec:sims}.  We see that the distributions are Gaussian, shown by the black curves, with $1\sigma$ marginalized errors obtained from the diagonal elements of the parameter-shift covariance matrix (described in Section~\ref{sec:param-shift-covmat}) given by the dashed vertical lines.  The solid vertical lines are the means of the distributions, and the dotted vertical lines are at zero.  We see no significant bias away from the expectation of a mean shift of zero.}
  \label{fig:ParamShiftBias2}
\end{figure*}

\begin{figure*}[ht]
  \centering
  \includegraphics[width=0.9\textwidth]{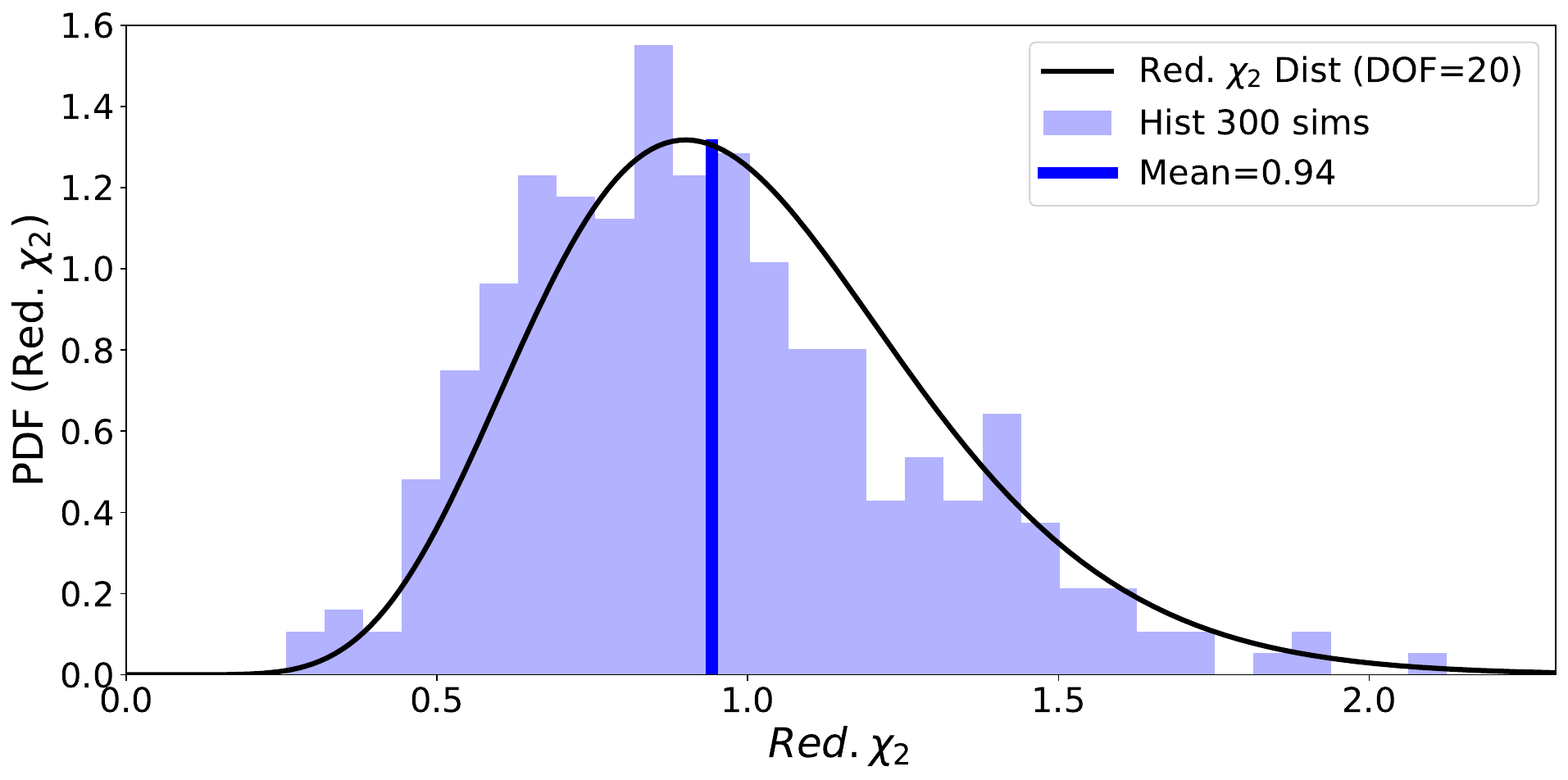}
  \caption{Shown is the histogram of reduced $\chi^2$ values from 300 simulated best-fit parameter shifts. Shown also is the mean reduced $\chi^2$ value for these simulations (vertical blue line).  The histogram matches the expected reduced $\chi^2$ distribution given 20 degrees of freedom (shown in black).}
  \label{fig:ParamShiftChi2}
\end{figure*}

\begin{figure*}[hb]
  \centering
  \includegraphics[width=0.9\textwidth]{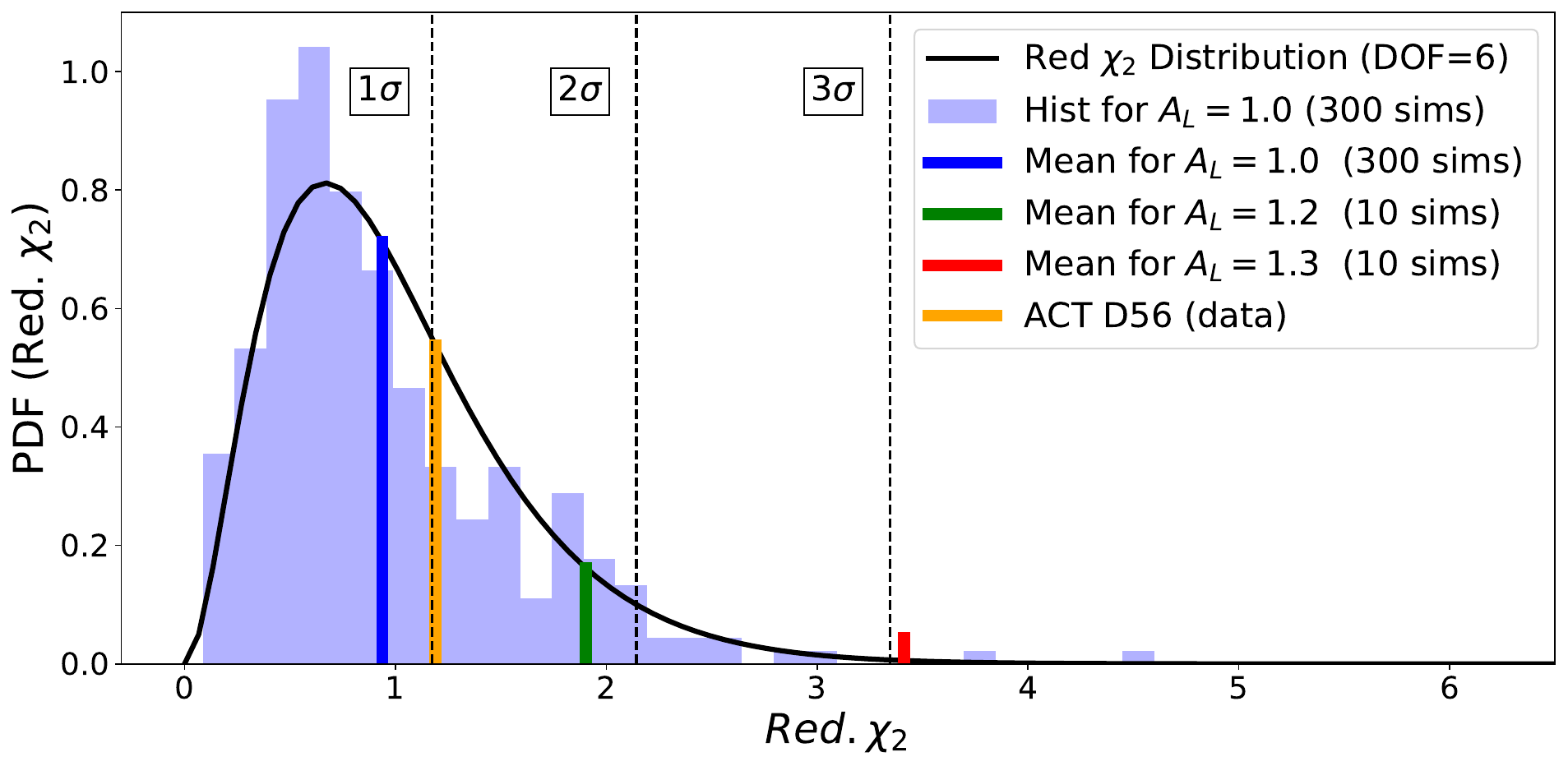}
  \caption{Shown is the histogram of reduced $\chi^2$ values from 300 simulated best-fit parameter shifts when marginalizing over foreground and calibration parameters and only considering the $\Lambda$CDM parameters. Shown also is the mean reduced $\chi^2$ value for these simulations (vertical blue line).  The vertical dashed black lines show the reduced $\chi^2$ values corresponding to $1\sigma$, $2\sigma$, and $3\sigma$ deviations from zero shift. These simulations are constructed with $A_L = 1.0$; we also show the mean reduced $\chi^2$ value calculated from 10 simulations constructed with $A_L = 1.2$ (vertical green line at $1.8\sigma$ deviation) and $A_L = 1.3$ (vertical red line at $3.1\sigma$) (see Section~\ref{sec:newPhysics}).   From this we see that if the CMB power spectrum contained a lensing-like signal that is $A_L = 1.2$ times that in $\Lambda$CDM, then our parameter-shift statistic (which is formulated assuming $\Lambda$CDM) would deviate at the $2\sigma$ level.  Similarly, it would deviate at the $3\sigma$ level for a lensing-like signal with $A_L = 1.3$. The vertical orange line gives the reduced $\chi^2$ value for the ACT D56 data.}
  \label{fig:AL}
\end{figure*}

As mentioned above, best-fit parameters inferred using the $action=2$ CosmoMC statistic have random scatter of about $2\%$ of the true best-fit parameter values. In order to reduce this scatter, we average many $action=2$ runs together for the same spectra.  To determine how many runs to average together to achieve stability in the best-fit parameters shifts, we use a simulation from simulation $set_3$ described in Section \ref{sec:sims} and run it through the pipeline described in Section \ref{sec:pipe} to generate a lensed and corresponding delensed spectra set. We then compute the difference in best-fit parameters for these spectra fifty times using the $action=2$ statistic, and show the convergence on the running average in Figure~\ref{fig:action=2Convergence}. 
For each iteration, we also compute the reduced $\chi^2$ for the average difference in parameters using the parameter-shift covariance matrix discussed in Section~\ref{sec:like}.  Averaging twenty versus fifty $action=2$ runs, changes the reduced $\chi^2$, and subsequent PTE, by $5\%$.  Thus, we conclude that the difference in best-fit parameters are sufficiently converged when averaged over twenty $action=2$ runs.

We also test the convergence of our parameter-shift covariance matrix by computing the running average and the running standard deviation of the parameter shifts derived from $N$ simulations.  Figure~\ref{fig:ParamShiftConvergence} shows that the mean converges to zero (red dots) and that the standard deviation of each parameter (blue stars) is stable after 200 simulations.  We use 300 simulations to generate the parameter-shift covariance matrix; thus we expect it to be well converged.  In general, using a finite number of simulations to estimate a covariance matrix can bias the resulting $\chi^2$ and PTE values. We use Equation 17 in \cite{Hartlap2007} to account for this bias by applying a multiplicative correction factor to the inverse of our covariance matrix; for our baseline analysis using 300 simulations, this correction is roughly equivalent to multiplying the parameter-shift covariance matrix by a factor of 1.07 (a $7\%$ increase in uncertainty). We apply this correction factor to all the parameter-shift PTE values presented in this analysis. 

We also verify that the parameter-shift covariance matrix is not biased.  Figure~\ref{fig:ParamShiftBias1} shows the histogram of best-fit parameters derived from each simulation realization in $set_3$.  Note that to include the impact of marginalizing over a Galactic dust prior and polarization foreground priors in simulations having no Galactic dust or polarized foregrounds, we impose the same prior widths as for the data but center the priors on zero. We find each histogram is well described by a Gaussian distribution centered at the input value used to generate these simulations. This figure also demonstrates that our best-fit-parameter measurements are not biased, both for lensed and delensed spectra.

Figure \ref{fig:ParamShiftBias2} shows the histogram of the best-fit parameter {\it{shifts}} derived from lensed versus delensed spectra. We find that these histograms each are also well described by a Gaussian distribution, this time centered at zero. Since both lensed and delensed spectra are derived from the same region of sky, we do not expect any mean difference in parameters derived from them.

We further check that the distribution of reduced $\chi^2$ values from the 300 simulations using this parameter-shift covariance matrix follows a $\chi^2$ distribution for 20 degrees of freedom, as we show in Figure~\ref{fig:ParamShiftChi2}. We also check that the distribution of reduced $\chi^2$ values follows a $\chi^2$ distribution when marginalizing over the foreground and calibration parameters (6 dof) as shown in Figure~\ref{fig:AL}.  Shown also is the mean reduced $\chi^2$ value for these simulations (vertical blue line). The vertical dashed black lines show the reduced $\chi^2$ values corresponding to $1\sigma$, $2\sigma$, and $3\sigma$ shifts.  These simulations are constructed with $A_L = 1.0$; we also show the mean reduced $\chi^2$ value calculated from 10 simulations constructed with $A_L = 1.2$ (vertical green line) and $A_L = 1.3$ (vertical red line) (see Section~\ref{sec:newPhysics}).  From this we see that if the CMB power spectrum contained a lensing-like signal that is $A_L = 1.2$ times that in $\Lambda$CDM, then our parameter-shift statistic (which is formulated assuming $\Lambda$CDM) would deviate at the $2\sigma$ level.  Similarly, it would deviate at the $3\sigma$ level for a lensing-like signal with $A_L = 1.3$.  The vertical orange line gives the reduced $\chi^2$ value for the ACT D56 data, which is consistent with $A_L = 1.0$.


\clearpage
\bibliographystyle{unsrturltrunc6}
\bibliography{delens}

\end{document}